\documentclass[11pt, a4paper]{article}
\pdfoutput=1
\usepackage{geometry}                % See geometry.pdf to learn the layout options. There are lots.
\usepackage{jcappub}
\usepackage{graphicx}
\usepackage{amssymb}
\usepackage{mathtools}
\usepackage{epstopdf}
\usepackage{verbatim}
\usepackage{color}
\usepackage{slashed}
\usepackage{bbold}
\usepackage{fullpage}
\usepackage{float}
\usepackage[title]{appendix}
\usepackage{amsfonts}
\usepackage{amssymb}
\usepackage{amsmath}
\usepackage{setspace}
\usepackage{epsfig}

\usepackage{amsthm}
\usepackage[usenames, dvipsnames]{xcolor}
\usepackage{tabularx}
\usepackage{empheq}
\usepackage{mathrsfs}

\definecolor{light-gray}{gray}{0.9}

\newcommand{\be}{\begin{eqnarray}}
\newcommand{\ee}{\end{eqnarray}}

\newcommand{\Lagr}{\mathcal{L}}
\newcommand{\bea}{\begin{align}}
\newcommand{\eea}{\end{align}}
\newcommand{\bmat}{\begin{pmatrix}}
\newcommand{\emat}{\end{pmatrix}}
\newcommand{\bal}{\begin{align}}
\newcommand{\eal}{\end{align}}

\definecolor{colorRTD}{rgb}{.2,.2,.7}

\newcommand{\w}{\omega}

\newcommand{\rhoDM }{\rho_{_{\text{DM}}}}

\newcommand{\mDM}{m_{_{\rm DM}}}

\newcommand{\mP}{m_{\rm Pl}}
\newcommand{\HII}{$\text{H}\scriptstyle{\text{II}}$}

\definecolor{myblue}{rgb}{.95, .95, 1}

\begin{document}

\subheader{}
\title{Premature Black Hole Death of Population III Stars by Dark Matter}

\author{Sebastian~A.~R.~Ellis}
\affiliation{D\'epartement de Physique Th\'eorique, Universit\'e de Gen\`eve,\\
24 quai Ernest Ansermet, 1211 Gen\`eve 4, Switzerland\\
\emph{and}}
\affiliation{Universit\'e Paris-Saclay, CNRS, CEA, Institut de Physique Th\'eorique, 91191, Gif-sur-Yvette, France}

\abstract{

Population III stars were the first generation of stars, formed in minihalos of roughly primordial element abundances, and therefore metal-free. They are thought to have formed at the cores of dense dark matter clouds. Interactions between baryons and dark matter can therefore have had an important impact on their evolution. In this paper we consider the capture of non- or weakly-annihilating dark matter by these early massive stars. In a wide region of parameter space, interactions of dark matter with baryons lead to premature death of the star as a black hole. We sketch how this modification of the standard evolutionary history of Population III stars might impact the epoch of reionisation, by modifying the amount of UV emission, the transition to Population II star formation, and the X-ray and radio emission from accretion onto the black hole remnants. Signals of massive black holes originating from Population III stars could be observed through gravitational waves from their mergers. Finally, the observation of pair-instability supernovae could effectively preclude premature black hole death across a wide range of parameter space, ranging in mass from $\mDM \sim 0.1~\text{GeV}$ to $\mDM \sim \mP$.
}
\maketitle

\section{Introduction}

The earliest stars are thought to have formed in the early universe at redshifts $z \leq 20$. Forming in minihalos with primordial element abundances, these stars were therefore roughly metal-free. As a result, this population of stars is referred to as Population III (Pop III) stars.
Throughout their life, Pop III stars might have had an important role in reionising the universe, producing regions of \HII~\cite{Barkana:2004vb,Furlanetto:2006jb,Schauer:2019ihk,Mirocha:2019gob,Mebane:2020jwl} through their emission of ultraviolet (UV) photons. As in life, so in death, with both the manner and remnants of Pop III star death also impacting this epoch. For certain mass ranges, these stars ended their lives as pair-instability supernovae (PISN) and ejected heavy elements, contributing to the enrichment of the intergalactic medium. The occurence of heavy element-forming supernovae affected the transition to formation of Population II (Pop II) stars, which had higher metallicity, and different UV emission properties. Pop III stars that ended their lives as black holes would lead to X-ray and radio emission due to accretion, further affecting the global 21-cm signal. These early stars therefore played an important role in an epoch we are only now beginning to probe~\cite{Furlanetto:2004nh,Bowman:2018yin,Mirocha:2019gob}. The search for these earliest stars, their death in PISN and their effect on the epoch of reionisation is a major goal for astronomy in the coming years~\cite{Mirocha:2019gob}.

Recently, the connection between dark matter (DM) and the era of reionisation has been of intense interest, primarily driven by EDGES and other experiments measuring the 21-cm signal. The EDGES collaboration has reported a strong global absorption signal~\cite{Bowman:2018yin}, which could be explained by enhanced DM-baryon interactions (see e.g.~\cite{Barkana:2018lgd,Munoz:2018pzp,Berlin:2018sjs,Barkana:2018qrx,Slatyer:2018aqg}). However, some concerns over the interpretation of the data have been raised, see e.g.~\cite{Hills:2018vyr,Bradley:2018eev,Sims:2019kro}. In this study, we consider how dark matter might have affected the death mechanism of Pop III stars. This modification to the standard Pop III evolutionary history could have indirect impacts on the global 21-cm signal, which we will discuss in general terms.

\begin{figure}[t]
\includegraphics[width=\textwidth]{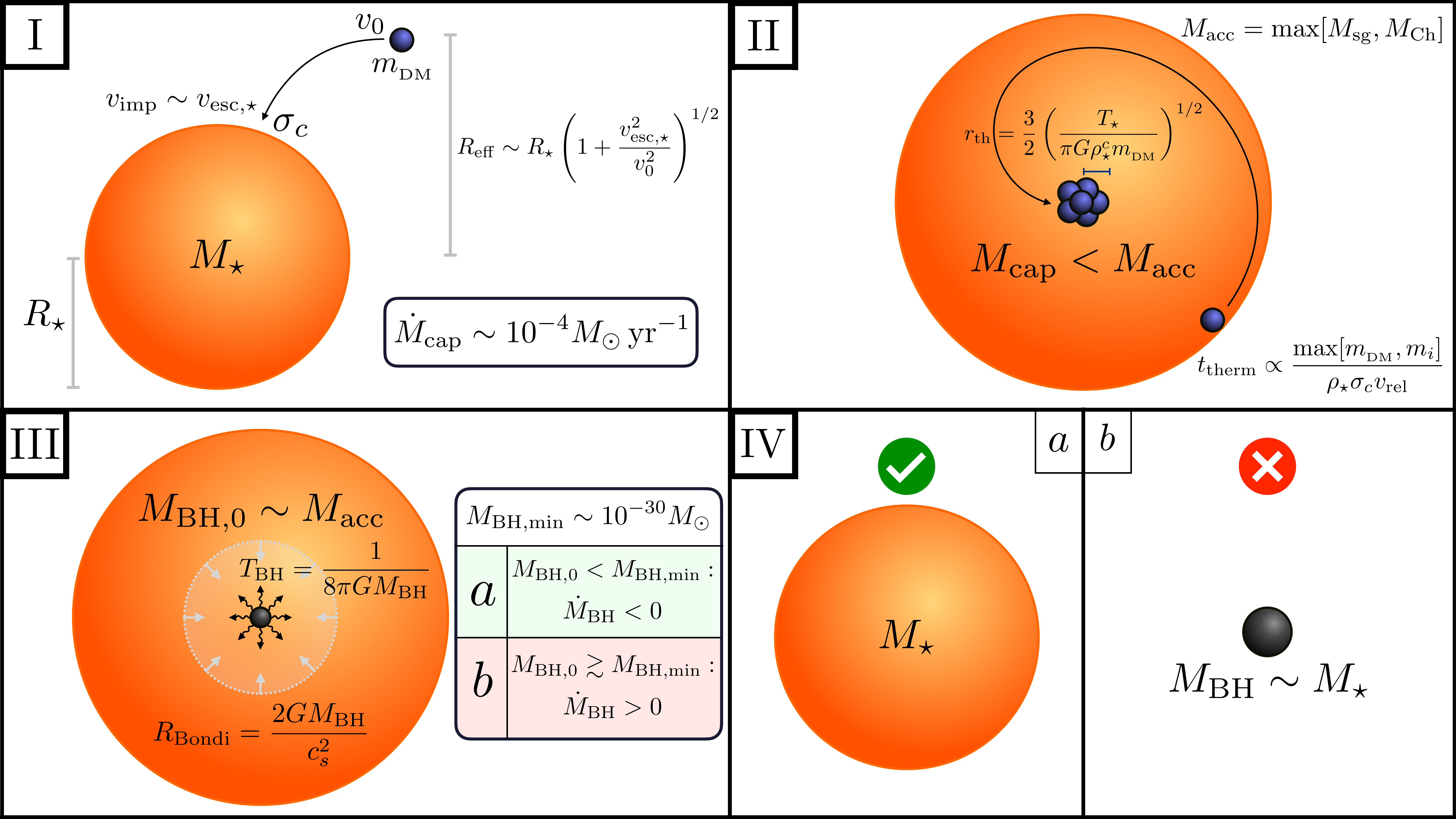}
\caption{Schematic showing how dark matter of mass $\mDM$ and cross-section $\sigma_c$ with baryonic matter can accumulate in a Pop III star (I and II), eventually collapsing into a black hole that can either accrete the surrounding matter to swallow the entire star, or evaporate (III and IV).}
\label{fig:Cartoon}
\end{figure}

If dark matter is weakly- or non-annihilating, as in models of asymmetric dark matter~\cite{Zurek:2013wia}, it is not depleted after being gravitationally captured by a star.\footnote{The study of dark matter capture on stars has a long history~\cite{Gould:1987ju,GOULD1990337,PhysRevD.40.3221,2008PhRvD..78l3510T,de_Lavallaz_2010,Frandsen_2010,Taoso_2010,Kouvaris:2010jy,Cumberbatch_2010,2012PhRvL.108f1301I,McDermott:2011jp,Kouvaris:2013kra,Baryakhtar:2017dbj,Bramante:2017xlb,Garani:2018kkd,PhysRevD.101.115021,Ilie:2020vec,Ilie:2020iup,Leane:2020wob,Acevedo:2020gro,Ilie:2021iyh,Gaidau:2021vyr,Leane:2021ihh}.} The accumulated dark matter will settle in the stellar core, where it will thermalise with the surrounding material. Eventually, enough dark matter will accumulate that a critical mass is reached, and pressure can no longer support the matter against gravitational collapse. A black hole (BH) will then form, which will either grow through accretion of the surrounding stellar matter and the continued infall of dark matter, or evaporate due to Hawking radiation. Simple arguments will demonstrate that in Pop III stars, the accretion rate of dark matter onto the BH exceeds the BH evaporation rate, even for Planck-mass dark matter. Therefore, given enough time, Pop III stars will eventually be consumed by a black hole of dark matter origin. The time required for consumption is the limiting factor for the formation and growth of black holes originating from light dark matter, and dark matter with either very small or very large baryonic interactions. Furthermore, evaporation of sub-GeV dark matter limits its retention and therefore collapse to a BH. For our benchmark Pop III star of Zero-Age Main Sequence (ZAMS) mass $M_{\rm Pop III} \sim 110\, M_\odot$ and radius $R_{\rm Pop III} \sim 5\,R_\odot$, dark matter below $m_\chi \lesssim 0.1\text{ GeV}$ will require more than the star's lifetime to accumulate the critical mass required to collapse into a BH. A cartoon depicting this mechanism is shown in Fig.~\ref{fig:Cartoon}.

Modifying the standard evolutionary paradigm of Pop III stars through premature black hole death affects reionisation in various ways. If Pop III stars' lives are severely curtailed, they will emit significantly fewer UV photons.
This could affect the timing of the global 21-cm signal, as UV emission is required to decouple the neutral hydrogen spin temperature from the background temperature through the Wouthuysen-Field effect~\cite{1952AJ.....57R..31W,1959ApJ...129..536F,1959ApJ...129..551F}. If the majority of Pop III stars end their lives as black holes as a result of dark matter accretion, the standard expectation of certain stars ending their lives as PISN would not be met. This can alter the transition from Pop III to Pop II star formation, further affecting the 21-cm signal. Finally, by increasing the number of black hole remnants from the first stars, an increase in the X-ray and radio emission from said remnants should be expected~\cite{1981MNRAS.194..639C,1995ApJ...438...40G,2001ApJ...561..496M,2007ApJ...662...53R,2008ApJ...680..829R}.\footnote{This emission is used to constrain the formation of primordial black holes, see e.g.~\cite{2017PhRvD..96h3524P,2017PhRvD..95d3534A} for CMB constraints. Recently this has been extended to the 21-cm signal~\cite{2021arXiv210811130Y}.} These emissions have opposite effects on the global signal. X-rays heat the intergalactic medium, reducing the size of the absorption feature, while radio emission could raise the background temperature above that of the cosmic microwave background (CMB), deepening the absorption feature. 

In this study, we will consider the dark matter mass and interaction range that can result in Pop III stars ending their lives prematurely as black holes, and comment on the potential impact. With the effectiveness of the mechanism firmly established, we will consider the details of how this impacts the era of reionisation in future work. We will also briefly comment on other implications of premature black hole death of Pop III stars, such as the possibility of constraining this mechanism through observation of a PISN, or through observation of gravitational wave events at LIGO/Virgo. 

The paper is organised as follows. In Section~\ref{sec:PopIII}, we give a brief overview of the standard histories of Pop III stars, e.g. masses, lifetimes and deaths. In Section~\ref{sec:Reionisation}, we review the global 21-cm signal. In Section~\ref{sec:Capture} we change tack and begin discussion of the role of dark matter, starting with the capture of dark matter onto the Pop III star. In this section, we compute the capture rate across a range of dark matter masses and interaction strengths, and compare our approach with others taken in the literature. In Section~\ref{sec:Thermalisation} we discuss the subsequent thermalisation of the gravitationally-bound dark matter, culminating with equilibration (and potentially evaporation) in the star's core. In Section~\ref{sec:Collapse} we discuss the conditions that must be satisfied for the accumulated dark matter to collapse into a black hole, and the time required for the process to take place. In Section~\ref{sec:BHEvol} we turn to discussing the subsequent evolution of the nascent black hole. In this section, we must consider the impact of potential non-spherical accretion onto the black hole due to the rotation of the Pop III star and the accreting dark matter. In Section~\ref{sec:Fate}, we summarise the fate of Pop III stars as a function of their mass, the density of dark matter in their host halo, the mass of dark matter, and the interaction cross-section. Having demonstrated that across a wide range of parameter space, Pop III stars can exprience premature black hole death, we examine possible consequences for the global 21-cm signal in Section~\ref{sec:DeathImpact}. We conclude in Section~\ref{sec:Consequences} by returning to the question of the possible consequences of premature black hole death.

\section{Standard Population III History and Cosmology}
\label{sec:PopIII}

Pop III stars are thought to have formed in metal-free minihalos of total mass $M \sim 10^6 M_\odot$.  Owing to the lack of metals at formation ($Z\lesssim 10^{-6} Z_\odot$), these stars are thought to have been very massive and luminous, due to the corresponding decreased efficiency of molecular hydrogen cooling (see e.g.~\cite{Bromm:1999du,Nakamura:2000ez,Abel:2001pr}). Furthermore, it is expected that these stars formed either in isolation, or in small clusters~\cite{Bromm:1999du,McKee:2007yx,Stacy:2013xwa}, depending on the degree to which they were influenced by other astrophysical sources. The very first generation of stars is expected to have formed in isolation, with a typical mass thought to be $M_\star \sim 100\,M_\odot$. In contrast, it is believed that the second generation of Pop III stars formed in small clusters and were lighter. This is expected due to processes that enhanced the electron fractions in the gas and therefore led to more efficient cooling and fragmentation (see e.g.~\cite{Greif:2006nr}). In what follows, we will primarily be concerned with the first generation of Pop III stars formed in isolation. 

The initial mass function (IMF) of Pop III stars is not well known. However, various groups have simulated the formation of these earliest stars and estimated the form of the IMF. Simulations tracking the formation of 110 Pop III stars yielded masses ranging from $10\,M_\odot$ to over $1000\,M_\odot$, peaked at around $100\,M_\odot$~\cite{Hirano:2013lba,Susa:2014moa}. The lower mass stars occurred in gas clouds formed by Hydrogen deuteride cooling as opposed to molecular hydrogen (H$_2$), and therefore evolved at lower temperatures. In minihalos supporting the formation of only one star, the average stellar mass found in Ref.~\cite{Susa:2014moa} was $M_\star \sim 125\,M_\odot$. Important factors determining the ultimate mass of a nascent Pop III star include the rotational velocity, the redshift of formation, and the minihalo virial mass~\cite{Hirano:2013lba}. 

Dark matter interactions with baryons could have had an impact on the formation of Pop III stars, thereby impacting their IMF. The impact of annihilating dark matter on Pop III formation was considered, for example, in Ref.~\cite{Stacy:2013xwa}. Meanwhile, large DM-baryon interactions could disrupt formation substantially, as shown in Ref.~\cite{2018MNRAS.480L..85H}. For $\mDM \lesssim 0.1\,\text{GeV}$ and $\sigma_c \gtrsim 10^{-21}\,\text{cm}^2$, the DM would cool the baryon gas, leading to a fragmentation mass scale of $M\lesssim 10\,M_\odot$, and therefore an IMF dominated by low-mass stars. Meanwhile, for $\mDM \gtrsim\text{GeV}$ and $\sigma_c/\mDM \gtrsim 10^{20}\,\text{cm}^2/\text{GeV}$, DM would heat the gas to above the virial temperature of the host halo, delaying formation until sufficiently large halos could form. We do not show the relevant regions in our figures, as their consistency with cosmological constraints rely on having velocity-dependent DM-SM interactions, introducing model-dependence we do not wish to consider here.

The typical lifetime of a Pop III star is on the order of $\text{few} \times 10^6~\text{yr}$, with only $\mathcal{O}(1)$ variability across stellar masses ranging from $40\,M_\odot$ to $915\,M_\odot$~\cite{2009ApJ...706.1184O}. As massive, hot and dense stars near the end of their lives, they can encounter a region of phase space known as the pair instability region. In this region, the density and temperature in the core are sufficiently high that $\text{e}^+\text{e}^-$ pairs can be produced despite the Boltzmann suppression. If these annihilate into neutrinos rather than producing photons, there is a decrease in radiation pressure supporting the star. This drop in radiation pressure can cause the core to partially collapse, resulting in runaway fusion culminating in a pair instability supernova (PISN), leaving no remnant behind~\cite{2002ApJ...567..532H,2003ApJ...591..288H}. 

The mass range of Pop III stars that end their lives as PISN depends on their rotational velocity. Large stellar rotational velocities reduce the mass scale at which a star can end its life in a PISN, since a critical factor is the mass of the helium core, which increases as a function of rotational velocity~\cite{2012A&A...542A.113Y}. For zero rotational velocity, the PISN mass range is roughly $120\,M_\odot \lesssim M_{\rm PISN} \lesssim 240\,M_\odot$, while a star rotating at 50\% of the corresponding Keplerian velocity of a test particle at the stellar surface would have a PISN mass range of roughly $90\,M_\odot \lesssim M_{\rm PISN} \lesssim 190\,M_\odot$~\cite{2012A&A...542A.113Y}. For a graphical representation of the dependence on rotational velocity, see Fig.~12 of Ref.~\cite{2012A&A...542A.113Y}.

Pop III stars with masses above the PISN range end their lives by directly collapsing to black holes. Meanwhile for masses below the PISN range, there is a region of parameter space where the core of the Pop III star skirts the pair instability region, and can undergo a pulsational PISN. For rapidly rotating stars above a certain threshold, chemical mixing becomes important~\cite{1987A&A...178..159M}, leading to modified evolution. This region can be loosely characterised as bounded by $30\,M_\odot \lesssim M_\star \lesssim 200\,M_\odot$ and $v_{\rm rot} \gtrsim 0.5\,v_{\rm K}$, where $v_{\rm K}$ is the Keplerian velocity. In this part of parameter space, Pop III stars may end their lives as PISN, pulsational PISN or be gamma-ray burst progenitors~\cite{2012A&A...542A.113Y}.

Since PISN typically result in no remnant there is a ``black hole mass gap'', where black holes would not have formed directly from stellar death. This mass gap lies roughly in the range $45\,M_\odot \lesssim M_{\rm BH} \lesssim 135\,M_\odot$~\cite{2021ApJ...910...30T}. Interestingly, LIGO and Virgo observed a gravitational wave signal in tension with this mass gap, consistent with the merger of $85^{+21}_{-14}\,M_\odot$ and $66^{+17}_{-18}\,M_\odot$ black holes~\cite{PhysRevLett.125.101102}. Such an event could have been the result of hierarchical merging of black hole remnants~\cite{PhysRevD.95.124046,Fragione:2020aki}. Dark matter-induced premature black hole death of Pop III stars is a mechanism which could populate this mass gap, and therefore potentially explain the observed event without the need for prior mergers.\,\footnote{Note that there have been other beyond the standard model explanations put forth to explain this event, e.g.~\cite{Sakstein:2020axg}.} 

Below this mass range, it is possible that other merger events detected by LIGO and Virgo~\cite{PhysRevX.9.031040} resulted from Pop III progenitors~\cite{2020MNRAS.498.3946K}. Since the mechanism we present here would apply to Pop III stars of all masses, detailed modelling of the expected merger rate would determine compatibility with the observed LIGO/Virgo event rate. Due to the significant uncertainty in the Pop III IMF, compatibility is plausible.

Ultimately, the Pop III origin of observables such as mergers of stellar remnants or the 21-cm signal discussed below could be debated. The direct observation of an individual Pop III star is unlikely with even the James Webb Space Telescope (JWST)~\cite{2006SSRv..123..485G,2013MNRAS.429.3658R}, as even a $1000\,M_\odot$ star at $z=30$ would have an absolute bolometric magnitude of $\sim 36$. The death of a Pop III star as a PISN, however, is an (almost) literal smoking gun demonstrating the past existence of an extremely massive, metal-poor star. The JWST might be able to detect the redshifted ultraviolet emission from such an explosion in its near-infrared camera (NIRCam) out to relatively high $z$~\cite{2018MNRAS.479.2202H}. 

\subsection{Standard Reionisation and Population III stars}
\label{sec:Reionisation}

The study of the cosmological impact of Pop III stars, and in particular the possibility that they played a role in reionisation, has a long history~\cite{1967AZh....44..295D,1977Ap&SS..48..145H,1984ApJ...277..445C}. Recent studies have focused on the potential impact of Pop III formation histories on the global 21-cm signal~\cite{Furlanetto:2006jb,Mirocha:2017xxz,Schauer:2019ihk,Mebane:2020jwl,Mirocha:2019gob}. One of the reasons for the intense interest is that Pop III stars are expected to have up to $10\times$ as much Lyman-Werner (LW) emission per baryon as Pop II stars~\cite{2002A&A...382...28S,Barkana:2004vb}. This LW emission can decouple the Hydrogen spin temperature from the CMB temperature through the Wouthuysen-Field effect~\cite{1952AJ.....57R..31W,1959ApJ...129..536F,1959ApJ...129..551F}, and therefore have a noticeable early impact on the global 21-cm absorption feature. Furthermore, black hole remnants of Pop III stars can be powerful X-ray and radio emitters as they accrete material, affecting both the timing and depth of the global 21-cm signal.

The size of the global 21-cm signal depends on the various relevant temperatures as
\begin{align}
\delta T_{\rm b} \propto 
\frac{T_s - T_{\rm rad}}{T_s} \ ,
\label{eq:deltaTb}
\end{align}
where $T_s$ is the spin temperature of neutral Hydrogen, given by
\begin{align}
T_s^{-1} = \frac{T_{\rm rad}^{-1} + x_\alpha T_\alpha^{-1} + x_c T_K^{-1}}{1+x_\alpha + x_c} \ .
\label{eq:Tspin}
\end{align}
The background temperature $T_{\rm rad}$ is usually the CMB temperature, $T_\alpha$ is the temperature of Lyman-$\alpha$ emission, and $T_K$ is the kinetic temperature of the gas. The parameter $x_\alpha$ is the radiative coupling coefficient quantifying the Wouthuysen-Field effect, and is therefore proportional to the Lyman-$\alpha$ flux, while $x_c$ is the collisional coupling coefficient. The shift in the temperature $\delta T_{\rm b}$ is therefore sensitive to the formation rate of Pop III stars and their lifetime through $x_\alpha$. It is further sensitive to the behaviour of possible black hole remnants through their impact on $T_{\rm rad}$ and $T_K$. If the remnants emit in radio, they can increase $T_{\rm rad}$, while if they emit large quantities of X-rays, this will increase $T_K$. Various studies on the impact of Pop III stars on reionisation have been conducted in recent years, e.g.~\cite{Mirocha:2017xxz,2018ApJ...868...63E,Schauer:2019ihk,Mebane:2020jwl}. We will return to the impact of premature black hole death on reionisation in Section~\ref{sec:DeathImpact}.

\section{Capture of Dark Matter on Population III Stars}
\label{sec:Capture}

Having briefly reviewed the standard evolution of Pop III stars and their importance for the epoch of reionisation, we now turn to the discussion involving dark matter. The first of a sequence of events culminating in the disruption of Pop III star evolution is the capture of dark matter onto the star. 

The capture of DM can be factorised into two parts: the flux on the star $\Phi_{\rm DM}$, and the probability that an incident particle will be captured $P_{\rm cap}$, such that the rate of mass capture can be written as
\begin{equation}
\dot{M}_{\rm cap} = \mDM \Phi_{\rm DM} P_{\rm cap} \ .
\label{eq:CaptureRate}
\end{equation}
The flux depends on the density and velocity distribution of the ambient DM, while the capture probability depends on the interaction strength between DM and nucleons, $\sigma_c$. We will consider short-range contact interactions, and therefore remain agnostic to the possible enhancement of capture rates due to long-range interactions~\cite{Gaidau:2021vyr}.

Population III stars, by virtue of forming in regions of very high DM density, will efficiently capture DM ranging in mass from sub-GeV to super-Planck, and with cross-sections varying across many orders of magnitude. As such, the capture calculation spans various regimes, each of which can be treated approximately analytically under certain assumptions. Throughout our analysis, when requiring knowledge of properties of the star, we use \texttt{MESA}~\cite{Paxton2011,Paxton2013,Paxton2015,Paxton2018,Paxton2019} output for the numerics.

The cross-section range probed by Pop III stars covers both the particle and fluid regimes, as defined by the Knudsen number (or alternatively the optical depth $\tau_\star$) written in terms of the mean free path of DM particles in the star, $\ell_{\rm DM} \equiv 
(n_\star \sigma_c)^{-1}$
\begin{align}
\text{Kn} \equiv \frac{\ell_{\rm DM}}{2R_\star } \equiv \frac{1}{\tau_\star} \sim 10^{-2} \left(\frac{5~ R_\odot}{R_\star} \right)\left(\frac{1.3\, \text{g cm}^{-3}}{\bar{\rho}_\star} \right)\left(\frac{2\times 10^{-34}\, \text{cm}^{2}}{\sigma_c} \right)\ ,
\end{align}
where in the second equality we have assumed that the DM has a contact interaction with the target nuclei, which we assume to be protons, as Pop III stars are expected to be made of roughly $75\%$ hydrogen. Note that the Knudsen number is effectively the inverse of the optical depth of DM scattering off the star. For small Knudsen numbers below $\text{Kn} \lesssim 10^{-2}$, corresponding to cross-sections $\sigma_c \gtrsim 10^{-34}\,\text{cm}^2$ for the choices above, the star is opaque to DM, and we can consider the flow to be in the fluid regime. Meanwhile, for $\text{Kn} \gtrsim 1$, corresponding to cross-sections $\sigma_c \lesssim 10^{-36}\,\text{cm}^2$ above, the flow is in the particle regime, as an incoming DM particle will on average experience less than one collision per traverse of the star. Between these limits is the ``multi-scatter'' regime, which has been developed and refined in Refs.~\cite{Bramante:2017xlb,Ilie:2020vec}. In this regime, the DM scatters an average of between 1 and 100 times as it traverses the star. 

The condition for gravitational capture of incident DM particles by the star can be estimated by requiring that the speed of the DM after having traversed a distance $x\sim R_\star$ be no greater than $v_{\rm esc, \star}$. The incoming DM has kinetic energy at the stellar surface of $E(R_\star) \sim \frac{1}{2}\mDM (v_{\rm esc,\star}^2 + v^2)$ due to gravitational focusing. In this expression, $v$ is the velocity at infinity of the DM, which should be pulled from the standard Maxwellian velocity distribution of DM in the Pop III's microhalo. The DM must therefore shed roughly $\Delta E \gtrsim \mDM v^2 / 2$ in order to be captured by the star. The maximal $v$ which can still be captured will be a function of the interaction strength between DM and the stellar material, as we will see below.

Let us first consider a single collision between a right-moving DM particle of mass $\mDM$ with initial velocity $\vec{v}$ and a left-moving target of mass $m_i$ moving with velocity $\vec{v}_i$. In the DM rest frame, assuming elastic collisions with zero impact parameter, we find that the DM's final velocity is given in terms of the reduced mass $\mu_{i,\rm DM}$ by
\begin{equation}
|\vec{v}^{\,\prime}| = \frac{2 |\vec{v}^b_i|}{\mDM} \mu_{ i,\rm DM} \ ,
\end{equation}
where the superscript $b$ denotes that this is the target velocity boosted to the DM rest frame. Boosting back to the stellar rest frame, we have
\begin{equation}
|\vec{v}^{\,\prime}|^2 = |\vec{v}|^2 - \frac{4 \mu_{i,\rm DM}}{\mDM}\left(|\vec{v}|\left(|\vec{v}|-|\vec{v}_i|\cos\theta \right) - \frac{\mu_{i,\rm DM}}{\mDM}|\vec{v}-\vec{v}_i|^2 \right) \ ,
\end{equation}
where $\theta$ is the angle between the initial DM and target velocities.
The average change in kinetic energy of an incoming DM particle as a result of a single collision is therefore
\begin{align}
\Delta E  = \frac{1}{2}\mDM(|\vec{v}^{\,\prime}|^2 - |\vec{v}|^2) \simeq - \frac{2 \mDM^2 m_i |\vec{v}|^2}{(\mDM + m_i)^2} + \frac{2 \mDM m_i^2 |\vec{v}_i|^2 }{(\mDM + m_i)^2} \ ,
\label{eq:deltaE}
\end{align}
where in the second equality we have taken the average over the collision angle $\cos\theta$. Gravitational focusing means that DM enters the star with $|\vec{v}| \gtrsim v_{\rm esc,\star}$, which for a Pop III star of mass $M_{\star} \sim 110 M_\odot$ and radius $R_{\star} \sim 5 R_\odot$ is $v_{\rm esc,\star} \sim 10^{-2}$. Meanwhile $|\vec{v}_i| \sim c_s(r)$, the stellar sound speed, which for a Pop III star ranges from $c_s(R_\star) \sim 10^{-4}$ to $c_s(0) \sim 5\times 10^{-3}$, with an average over the star of $\bar{c}_s \sim 3 \times 10^{-3}$. 

\subsection{The Fluid Limit}

The change in DM energy per distance travelled between scattering events $\ell_{\rm DM}$ is simply
\begin{equation}
\frac{\Delta E}{\ell_{\rm DM}} = n_i \sigma_c \Delta E \ ,
\end{equation}
where $n_i$ is the number density of target particles, and can be related to the density by $n_i = \rho_i / m_i$.\footnote{Since a typical Pop III star is roughly $75\%$ Hydrogen and $25\%$ Helium, we will take $m_i \sim m_\text{H}$, and $\rho_i \sim \rho_\star$. The impact of collisions on Helium on the capture rate has been considered in Ref.~\cite{Ilie:2021iyh}, and found to enhance the capture rate.} In the fluid limit, $\ell_{\rm DM} \ll R_\star$, and $\text{Kn} \to 0$, so that we may write this equation as
\begin{align}
\frac{dE}{dx} \sim \frac{\rho_i}{m_i} \sigma_{c} \Delta E \ .
\label{eq:dEdx}
\end{align}
In the large $\mDM |\vec{v}|^2$ limit, the first term on the RHS of Eq.~\eqref{eq:deltaE} dominates. Since this term is proportional to $E$, the incoming DM loses energy exponentially fast as it penetrates the star. In the small $\mDM |\vec{v}|^2$ limit, the energy loss is linear as the second term in Eq.~\eqref{eq:deltaE}, which is independent of $E$, dominates. From the above, we can estimate the maximal velocity $v_{\rm max}$ the DM can have far from the star
\begin{align}
v_{\rm max} &\sim v_{\rm esc,\star}\left(\exp\left(\frac{2 R_\star \bar{\rho}_\star \sigma_{c}}{\mDM}\right) -1 \right)^{1/2} \ , ~~~~\mDM |\vec{v}|^2 \gg m_i |\vec{v}_i|^2 \ , 
\label{eq:vMax1}\\
v_{\rm max} &\sim v_i  \frac{(2m_i \bar{\rho}_\star R_\star\sigma_{c})^{1/2}}{\mDM + m_i} \ , \hspace{3cm}\mDM |\vec{v}|^2 \ll m_i |\vec{v}_i|^2 \ .
\label{eq:vMax2}
\end{align}
such that for $v \lesssim v_{\rm max}$, the DM will be gravitationally captured after a stellar transit.

Thus, when we are in the fluid limit,  the capture probability $P_{\rm cap}$ in Eq.~\eqref{eq:CaptureRate} is simply
\begin{equation}
P_{\rm cap} = \Theta(v_{\rm max} - v) \ ,
\end{equation}
where $\Theta(x)$ is the Heaviside step function.

\subsection{The Particle Limit}

In the particle limit, the average total energy lost as a result of $n_{\rm coll}$ collisions is simply
\begin{equation}
\Delta E_{\rm tot} = \frac{1}{2}\mDM(v_f^2 - v_0^2) \sim \sum_I^{n_{\rm coll}} \Delta E_I \ , 
\end{equation}
where $v_f$ is the final velocity of the DM after $n_{\rm coll}$ collisions and $v_0$ is the velocity upon entering the star. We may now use Eq.~\eqref{eq:deltaE} written in the following suggestive form 
\begin{equation}
\Delta E_I \simeq -\frac{1}{2}\mDM v_{I-1}^2 \left( \frac{4\mDM m_i }{(\mDM + m_i)^2} + \frac{v_i^2}{v_{I-1}^2}\frac{4m_i^2}{(\mDM + m_i)^2} \right) \ ,
\label{eq:deltaEi}
\end{equation}
where we have approximated $|\vec{v}_x|\sim v_x$ for both the DM and target velocities. This form is chosen for simple comparison with the literature~\cite{Bramante:2017xlb,Ilie:2020vec}, where the variable $\beta_+ \equiv \frac{4\mDM m_i }{(\mDM + m_i)^2}$ is introduced.

The total energy loss after $n_{\rm coll}$ collisions can be written in concise forms in the relevant limits of the ratio $\mDM v_{\rm esc,\star}^2 / m_i v_i^2$. We recall from the discussion below Eq.~\eqref{eq:deltaE} that the hierarchy between $v$ and $v_i$ in the star is on average only $v_{\rm esc,\star}/v_i\sim3$ for a Pop III star of $M_\star \sim 110 M_\odot$ and radius $R_\star \sim 5 R_\odot$. Thus, for $\mDM \gg (v_{\rm esc,\star}/v_i)^2 m_i \sim 10\, m_i$, we may write
\begin{equation}
\Delta E_{\rm tot} = \frac{1}{2}\mDM v_0^2\left[\left(1-\frac{\beta_+}{2} \right)^{n_{\rm coll}} - 1 \right] \ .
\end{equation}
In the above, we note that due to gravitation, $v_0^2 = v_{\rm esc,\star}^2 + v^2$, where $v$ is drawn from the Maxwellian velocity distribution.
However, when $\mDM \ll (v_{\rm esc,\star}/v_i)^2 m_i$, we can see from Eq.~\eqref{eq:deltaEi} that it is the second term that dominates, and we are in the linear energy loss regime. Then we find
\begin{equation}
\Delta E_{\rm tot} = - n_{\rm coll} \mDM v_i^2 \frac{m_i^2}{(\mDM + m_i)^2} \ .
\end{equation}
In the particle regime, the maximum velocity of an incoming DM particle can have to be captured is
\begin{align}
v_{\rm max} &\sim v_{\rm esc,\star}\left(\left(1-\frac{\beta_+}{2} \right)^{-n_{\rm coll}} -1 \right)^{1/2} \ , ~~~~\mDM |\vec{v}|^2 \gg m_i |\vec{v}_i|^2 \ , \\
v_{\rm max} &\sim v_i \sqrt{2 n_{\rm coll}} \frac{m_i}{\mDM + m_i} \ , \hspace{2.4cm}\mDM |\vec{v}|^2 \ll m_i |\vec{v}_i|^2 \ .
\label{eq:vMaxParticle}
\end{align}
These expressions should be compared against Eqs.~\eqref{eq:vMax1},~\eqref{eq:vMax2}.

The average number of collisions experienced by the DM as it travels a distance $R_\star$ through the star is $\langle n_{\rm coll} \rangle \sim R_\star \bar{\rho}_\star \sigma_c / m_i$. Inserting this into the above in the place of $n_{\rm coll}$, and comparing with Eqs.~\eqref{eq:vMax1},~\eqref{eq:vMax2} we find that in both the large and small $\mDM$ limits, $v_{\rm max}$ is \textit{the same} in both the particle and fluid approach to the calculation. In the small cross-section $\sigma_c$ limit, corresponding to large $\text{Kn}$, approximating $n_{\rm coll}$ by $\langle n_{\rm coll} \rangle$ is appropriate, since the number of collisions is Poisson-distributed, and in the limit where $\sigma_c \to 0$, $\langle n_{\rm coll} \rangle\to 0$, and $P(n_{\rm coll} = \langle n_{\rm coll} \rangle) \to 1$. Thus, in the regime where $\text{Kn} \gtrsim 100$, the results of the fluid regime \emph{can} be used at the cost of being accurate to within a few percent. In the intermediate regime for $10^{-2} \lesssim \text{Kn} \lesssim 100$, one should use the multi-scatter sum over different $n_{\rm coll}$, weighted by the Poisson probability distribution function, given the expectation value. However, this regime covers only 4 orders of magnitude in cross-section probed. For the reference Pop III star of $M_\star \sim 110 M_\odot$ and radius $R_\star \sim 5 R_\odot$, this band is at $10^{-38} \lesssim \sigma_c/\text{cm}^2 \lesssim 10^{-34}$. As we will see in Section~\ref{sec:Fate}, Pop III stars can probe parameter space both at much larger and smaller cross-sections than this range. Since the fluid regime expressions are valid both deep in the fluid and particle regimes, we will use them for all the parameter space we consider. The cost of this approach is that the parameter space in the cross-section band described above will be subject to $\mathcal{O}(1)$ uncertainties. Since there are similar or larger uncertainties surrounding Pop III stars in general, we will accept this as tolerable error. Note also that the entire discussion thus far has been on a per-particle basis. However, the DM densities in the vicinities of Pop III stars are typically very large, $\rhoDM \gtrsim 10^{8}\, \text{GeV/cm}^3$. The number of DM particles transiting and interacting with the star every second is therefore large, even for $\mDM \sim \mP$ and cross-sections smaller than $10^{-34}~\text{cm}^2$. When averaging the above results over the large number of DM particles colliding with the star, it follows that the fluid calculation should function well in the whole parameter space we will be probing.

Thus, to conclude this subsection, we remark that in the particle regime we use Eqs.~\eqref{eq:vMax1},~\eqref{eq:vMax2} and therefore, as in the fluid regime, we can write
\begin{equation}
P_{\rm cap} = \Theta(v_{\rm max} - v) \ ,
\end{equation}
to compute the capture rate.

\subsection{Dark Matter Flux on the Star}

\begin{figure}[t]
\centering
\includegraphics[scale=1]{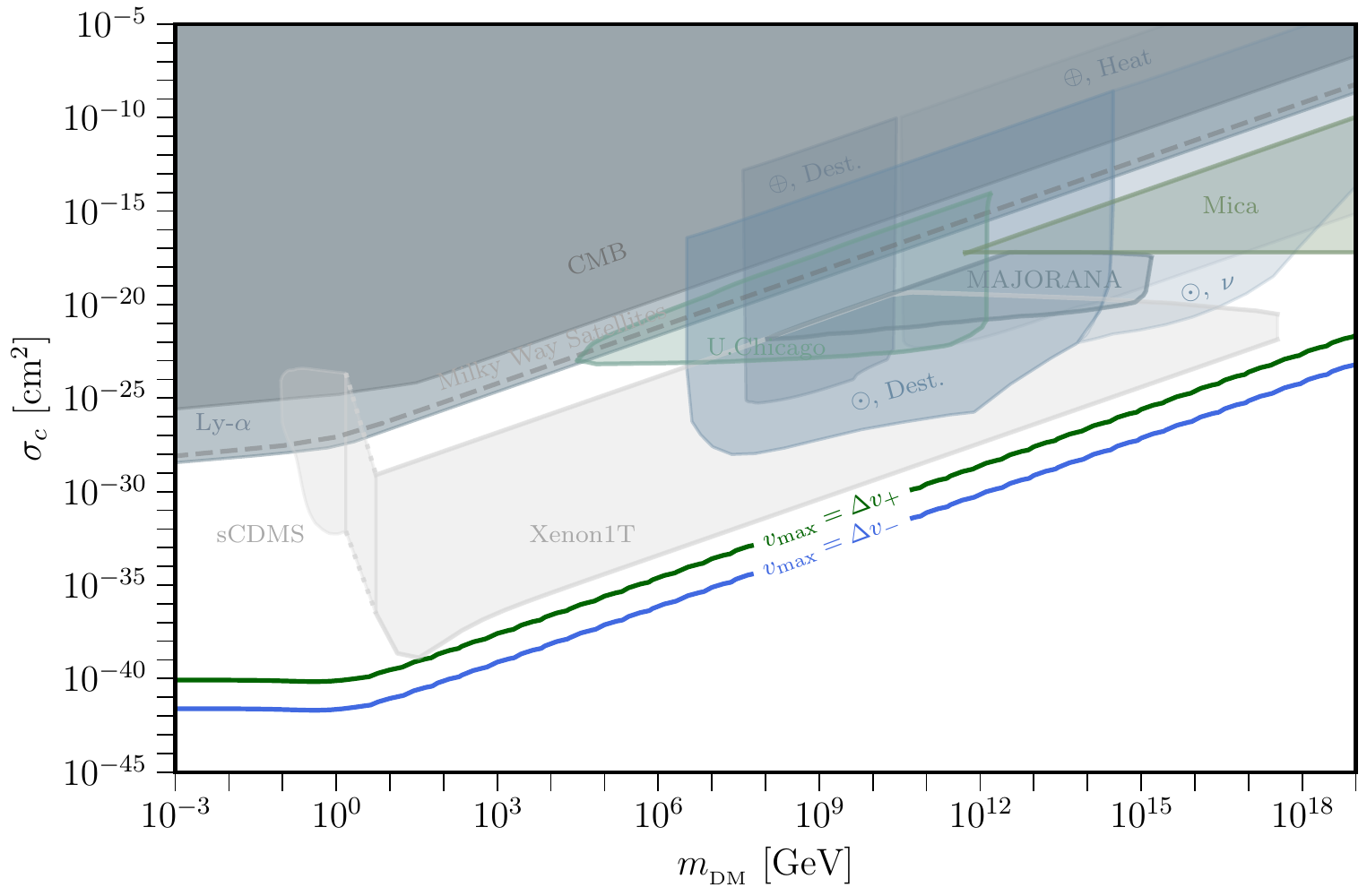}
\caption{The maximum velocity $v_{\rm max}$ regimes relative to $\Delta v_{\pm} \equiv v_{\rm esc,~MH} \pm v_\star$ are shown. The maximum velocity for capture is defined by Eq.~\eqref{eq:vMax1} in the large DM mass limit, and Eq.~\eqref{eq:vMax2} in the small DM mass limit. 
Above the green contour, where $v_{\rm max} \geq \Delta v_+$, the capture rate is approximately given by Eq.~\eqref{eq:captureBigvMax}, while below the blue contour, where $v_{\rm max} \leq \Delta v_-$, the capture rate is approximately given by Eq.~\eqref{eq:captureSmallvMax}. 
The change in scaling at $\mDM \lesssim 10~\text{GeV}$ is due to the assumption that the DM is scattering off Hydrogen moving with an average sound speed $c_s \sim v_{\rm esc,\star}/3$, such that Eq.~\eqref{eq:vMax2} applies instead of Eq.~\eqref{eq:vMax1}. See the main text for further details. Existing constraints on the parameter space from cosmology, milky way satellites, direct detection, and black hole formation in the Sun and Earth are also shown as different grey shaded regions~\cite{XENON:2017vdw,Xu:2018efh,Boddy:2018kfv,Digman:2019wdm,Nadler:2019zrb,Cappiello:2020lbk,SuperCDMS:2020aus,Clark:2020mna,Acevedo:2020gro,Buen-Abad:2021mvc}. Additional constraints, e.g.~\cite{CRESST:2017ues,CRESST:2019axx,CRESST:2019jnq,EDELWEISS:2019vjv,Erickcek:2007jv,Mahdawi:2018euy,McCammon:2002gb,Bringmann:2018cvk} are not shown for clarity, as they tend to overlap other constrained regions.}
\label{fig:velRegime}
\end{figure}

The flux of DM particles on the star is given by
\begin{align}
\Phi_{\rm DM} = \pi R_\star^2 n_{\rm DM} \int d^3 v f(\vec{v}) v\left(1 + \frac{v_{\rm esc,\star}^2}{v^2} \right) \ ,
\label{eq:DMFlux}
\end{align}
where the velocity distribution $f(\vec{v})$ is the usual Maxwellian, truncated to exclude velocities greater than the minihalo escape velocity $v_{\rm esc,~MH}$, and boosted to the stellar rest frame
\begin{align}
f(\vec{v}) &= \frac{\exp\left(-\frac{|\vec{v} + \vec{v}_\star|^2}{v_0^2} \right) }{\mathcal{N}} \Theta(v_{\rm esc,~MH} - |\vec{v} + \vec{v}_\star|) \ , \\
\mathcal{N} &= \pi^{3/2} v_0^3 \left(\text{erf}\left(\frac{v_{\rm esc,~MH}}{v_0} \right) - \frac{2 v_{\rm esc,~MH}}{\sqrt{\pi}v_0} \exp\left(- \frac{v_{\rm esc,~MH}^2}{v_0^2} \right)\right) \ ,
\end{align}
where the normalisation factor $\mathcal{N}$ is introduced to ensure that $4\pi \int_0^\infty  dv v^2 f(v) = 1$. The typical dispersion velocity in a Pop III star-forming minihalo is $\sigma_v \sim 5 \text{ km/s}$ (see e.g.~\cite{Stacy:2013xwa}), and $\sigma_v = \sqrt{3/2} v_0$, so that a typical value is $v_0 \sim 4 \text{ km/s}$. This is corroborated if one assumes an NFW profile for the minihalo formed at redshift $z\sim 15$, with concentration parameter $c\sim 2$~\cite{Ludlow_2016} and uses the virial theorem to compute the dispersion velocity of a $M \sim 10^6 M_\odot$ minihalo. The corresponding escape velocity for the minihalo is $v_{\rm esc,~MH} \sim 6~\text{km/s}$. 

Care must be taken when integrating Eq.~\eqref{eq:DMFlux}. Depending on the regime of $v_{\rm max}$, the polar angle defined by $\vec{v} \cdot \vec{v}_\star = v v_\star \cos\theta$ can be restricted, as DM moving too quickly at angles away from the star will never collide with the star. For notational ease, we define $\Delta v_\pm  \equiv v_{\rm esc,~MH} \pm v_\star$. The three cases are given here for concreteness:
\begin{itemize}
  \item For $v_{\rm max} \lesssim \Delta v_{-}$, there is no restriction on the polar angle, and $\int d^3 v \to 4\pi \int_0^{v_{\rm max}} dv\,v^2 $. 
  
  \item If $\Delta v_{-} \lesssim v_{\rm max} \lesssim \Delta v_{+}$, the integral must be performed over two regions. In the first, bounded above by $\Delta v_{-}$, there is no angular restriction, and $\int d^3 v \to 4\pi \int_0^{\Delta v_{-}} dv\,v^2 $. In the second, for $ \Delta v_{-} \lesssim v \lesssim v_{\rm max}$, the maximum scattering angle is defined by
\begin{equation}
\cos\theta_- \equiv \frac{v_{\rm esc,~MH}^2 - v_\star^2 - v^2}{2v_\star v} \ .
\end{equation}
In this region, 
$
\int d^3v \to 2\pi \int_{\Delta v_-}^{v_{\rm max}} dv\, v^2 \int_{-1}^{\cos\theta_-} d\cos\theta 
$.
\item For $\Delta v_{+} \lesssim v_{\rm max}$, there are three regions of integration. The first considers velocities smaller than $\Delta v_-$, for which the integral is performed as $\int d^3 v \to 4\pi \int_0^{\Delta v_{-}} dv\,v^2 $, with no truncation on the allowed scattering angle. The second region corresponds to velocities between $\Delta v_-$ and $\Delta v_+$, and is
$\int d^3v \to 2\pi \int_{\Delta v_-}^{\Delta v_+} dv\, v^2 \int_{-1}^{\cos\theta_-} d\cos\theta 
$. Clearly, velocities between these two values must have their angular integral truncated at the maximum scattering angle. Finally the third region, where the velocity integral should be performed from $\Delta v_+$ to $v_{\rm max}$ is zero, as these DM particles move away from the star and escape the halo.
\end{itemize}
Full analytic expressions for the integrals in these three regions can be found in Appendix C of Ref.~\cite{PhysRevD.101.115021}. 

\subsection{Approximate Capture Rates}

Having determined that in the totality of the parameter space we consider, the fluid approach works well for obtaining the capture probability, and then calculating the flux of DM incident on the star, we are now in a position to give approximate expressions for the capture rate in the various regimes we consider. The majority of the parameter space falls into either $v_{\rm max} \leq \Delta v_-$ or $v_{\rm max} \geq \Delta v_+$, as seen in Fig.~\ref{fig:velRegime}. In the latter limit, the accretion rate takes on the simple form
\begin{equation}
\dot{M}_{\rm acc} \sim \frac{2\pi G M_\star R_\star \rhoDM}{v_0} \ ,
\label{eq:captureBigvMax}
\end{equation}
which exhibits no dependence on the DM-SM interaction cross-section, and no dependence on the DM mass if the DM energy density is fixed. 
In this regime, the DM-SM interaction is sufficiently strong that all DM that encounters the star will be captured. Conversely, when $v_{\rm max} \lesssim \Delta v_-$, we find
\begin{align}
\dot{M}_{\rm acc} \sim \frac{G M_\star^2}{R_\star v_0^3} \rhoDM \sigma_{c} \times \begin{cases} 
\frac{G M_\star}{R_\star} \frac{1}{\mDM} \ , ~~ \mDM v_{\rm esc,\star}^2 \gg m_i|\vec{v}_i|^2 \ , \\
\frac{T_\star}{m_i^2} \ \ , ~~ \mDM v_{\rm esc,\star}^2 \ll m_i|\vec{v}_i|^2 \ ,
\end{cases}
\label{eq:captureSmallvMax}
\end{align}
which exhibits the expected dependence of the accretion rate on the DM-SM interaction strength. Of note however is the saturation at small $\mDM \ll m_i$, where the leading contribution to the capture rate is independent of the DM mass for a fixed DM energy density, and instead depends on the temperature of the stellar interior $T_\star$, which appears because $v_i \sim c_s \sim \sqrt{T_\star/m_i}$ in Eq.~\eqref{eq:vMax2}. This is expected from a thermodynamic perspective, since for small DM masses, the virial ``temperature'' of the DM bath is lower than that of the SM stellar bath. 

The total amount of DM captured on the star depends crucially on the precise value of $\rhoDM$ in the vicinity of the Pop III star. This quantity is not well known, although studies have suggested that at the effective capture radius $R_{\rm eff} \sim R_\star (v_{\rm esc}/v_0) \sim 10^{-4} \text{ pc}$, an adiabatically contracted NFW halo would have a DM density of roughly $\rhoDM \sim 10^{11} \,\text{GeV/cm}^3$, while if adiabatic contraction of the halo did not occur, one might expect a DM density of $\rhoDM \sim 10^8\,\text{GeV/cm}^3$~\cite{Freese:2008hb,Ilie:2020iup}. We will take the adiabatically contracted result as our benchmark, and comment on the variation due to changing $\rhoDM$ in Section~\ref{sec:Fate}.

\section{Thermalisation into a Dark Matter Core}
\label{sec:Thermalisation}
The above accretion calculation required that the DM shed sufficient energy for its velocity upon exiting the star to be below the escape velocity of the star. This is sufficient to demonstrate that the DM is gravitationally bound to the star, but it is clear that a DM particle exiting the star with speed $v = v_{\rm esc,\star}(1-\epsilon)$ with $\epsilon \ll 1$ will have an orbit that takes it far outside of the star. Thus, we must account for the time required for the DM orbits to converge to within the star in a first step, known as the ``first thermalisation time'', and then for the DM to settle into a thermal sphere at the core of the star, the ``second thermalisation time''.

\subsection{First thermalisation time}

The kinetic energy of the DM upon exiting the star is $E = \frac{1}{2} \mDM (v_{\rm esc,\star}(1-\epsilon))^2$. For simplicity, we will assume that all captured DM will pass through the core of the star. For DM transiting the star many times on its way to forming an orbit fully enclosed in the star, this is not necessarily a bad assumption. If the DM initially entered the star at an oblique angle, its orbit will be deflected as it transits, such that the exit angle will be more oblique.\footnote{Most DM entry angles will be oblique due to gravitational focusing.} This process repeats with each transit, with the DM completing ever-smaller quasi-elliptical orbits that have a periapsis closer to the stellar core each time, as the semi-major axis of the orbit shrinks. On average, therefore, DM incident at an oblique angle will transit more than $2R_\star$ through the star, thereby leading to greater energy loss than when assuming zero impact parameter DM-star collisions with repeated core transits.

If the DM transits the centre of the star multiple times as it ``orbits'' the star, then its orbit can be well approximated by a 1-dimensional problem analogous to a pendulum that experiences aerodynamical drag only in a region of width $2R_\star$. From energy conservation, the apex of the DM trajectory $r_{\rm max}$ at which the DM velocity is zero is found from
\begin{equation}
E_{\rm exit} - \frac{G M_\star \mDM}{R_\star} = - \frac{G M_\star \mDM}{r_{\rm max}} \ ,
\label{eq:ECons}
\end{equation}
such that
\begin{equation}
r_{\rm max} = \frac{G M_\star \mDM R_\star}{G M_\star \mDM - E_{\rm exit}R_\star} \ .
\end{equation}
We wish to solve $d r_{\rm max}/dt$, subject to appropriate boundary conditions. After the first transit, the DM exits with a velocity we will define as $v_{\text{exit},i} = v_{\rm esc,\star}(1-\epsilon)$. Thus, recalling that $v_{\rm esc,\star} = \sqrt{2 G M_\star/R_\star}$, the initial boundary condition is
\begin{equation}
r_{\text{max},i} = \frac{R_\star}{(2-\epsilon)\epsilon} \ .
\end{equation}
The final boundary condition is $r_{\text{max},f} = R_\star$, such that the orbit is fully contained in the star. The only quantity in $r_{\rm max}$ which changes as a function of time is $E_{\rm exit}$, so that
\begin{equation}
\frac{d r_{\rm max}}{dt} = \frac{r_{\rm max}^2}{G M_\star \mDM} \frac{d E_{\rm exit}}{dt}\ .
\end{equation}
In this expression, we can approximate $dE_{\rm exit}/dt$ as $\langle \Delta E \rangle / \Delta t$, where 
$\langle \Delta E \rangle$ is the average energy lost per transit, and $\Delta t$ is the average time between transits. The former can be computed from Eq.~\eqref{eq:dEdx} and Eq.~\eqref{eq:deltaE}. Assuming a transit distance of $2R_\star$, this gives
\begin{equation}
\langle \Delta E \rangle \simeq - R_\star \bar{\rho}_\star \sigma_c \times \begin{cases}  \frac{2 E_{\rm exit}}{\mDM} \ , ~ ~~~\mDM v^2 \gg m_i v_i^2 \ ,\\
 \frac{\mDM v_i^2}{m_i}\ , ~\mDM v^2 \ll m_i v_i^2 \ .
 \label{eq:avgdE}
\end{cases}
\end{equation}
Meanwhile $\Delta t$ can be computed by solving for the time taken to reach the apex and return to the stellar surface, approximating the average gravitational acceleration between $R_\star$ and $r_{\rm max}$ as 
\begin{equation}
g \sim \frac{\int_R^{r_{\rm max}} -\frac{G M_\star}{r^2} dr}{\int_R^{r_{\rm max}} dr} = - \frac{G M_\star}{R_\star r_{\rm max}} \ ,
\end{equation}
so that to find $\Delta t$, we must solve
\begin{equation}
\ddot{r}(t) = g \ ,~~ r(0) = R_\star,~ \dot{r}(0) = \sqrt{2E_{\rm exit}/\mDM} \ .
\end{equation}
Given that $E_{\rm exit} = G M_\star \mDM \left( \frac{1}{R_\star}- \frac{1}{r_{\rm max}} \right)$ from Eq.~\eqref{eq:ECons}, we find that 
\begin{equation}
\Delta t = \frac{2 R_\star r_{\rm max}}{G M_\star}\sqrt{2 G M_\star \left( \frac{1}{R_\star}- \frac{1}{r_{\rm max}} \right)} \ ,
\end{equation}
and therefore arrive at
\begin{align}
t_{\rm therm, 1} &= \int_{R_\star/(2-\epsilon)\epsilon}^{R_\star} d r_{\rm max} \frac{G M_\star \mDM \Delta t (r_{\rm max})}{\langle \Delta E (r_{\rm max}) \rangle }  \\
&\simeq \frac{2}{v_{\rm esc,\star} \bar{\rho} \sigma_c}\begin{cases}
\mDM\, \text{artanh}(1-\epsilon)\ , ~~~~~~~~~~~~~~~~~~ \mDM v_{\rm esc,\star}^2 \gg m_i v_i^2 \\
 m_i \frac{v_{\rm esc,\star}^2}{v_i^2}(\epsilon-1 + \text{artanh}(1-\epsilon) ) \ , ~~~ \mDM v_{\rm esc,\star}^2 \ll m_i v_i^2 
 \end{cases} \ .
 \label{eq:tTherm1}
\end{align}
 In both the small and large DM mass limits, this thermalisation time exhibits the same scaling as the time required to capture DM, as expected since the process governing these is the same, namely energy loss through collisions with the stellar material. Of note is that this scaling with $\mDM$ and $\sigma_c$ is the same as in Ref.~\cite{Kouvaris:2010jy}, but not as in Ref.~\cite{Acevedo:2020gro}.

 \subsubsection{First Thermalisation: Elliptical Orbits}

Our above calculation of the first thermalisation time differs from a recent calculation appearing in the literature, Ref.~\cite{Acevedo:2020gro}. In particular, their result scales as $\mDM^{3/2}$ in the large DM mass limit, unlike the $\mDM$ scaling we recover here. With $\mDM^{3/2}$ scaling, first thermalisation isochrones eventually intersect isochrones of single-transit thermalisation, an inconsistency. Single-transit thermalisation times can be obtained by imposing that the DM velocity $v \sim 0$ after transiting an order $R_\star$ distance when obtaining $v_{\rm max}$. As a cross-check, single-transit capture corresponds to $\epsilon = 1$, in which case $t_{\rm therm,1}=0$ above as expected.

 The approach taken in Ref.~\cite{Acevedo:2020gro} assumes 2-dimensional elliptic orbits for the captured DM particles, contrary to the oscillatory 1-dimensional orbits considered above.
 Elliptic orbits can be defined by two parameters, the semi-major axis $a$ and the eccentricity $e$. The energy of an elliptic orbit only depends on the semi-major axis, and is
 \begin{equation}
 E = -\frac{G M_\star \mDM}{2 a} \ ,
 \label{eq:orbitalEnergy}
 \end{equation}
 such that the orbital period can be written in terms of the semi-major axis
 \begin{equation}
  \tau = 2\pi \sqrt{\frac{a^3}{G M_\star}} \ .
 \end{equation}

 We may repeat the analysis done for the 1-dimensional simplified case above for the 2-dimensional elliptic orbit case. From Eq.~\eqref{eq:orbitalEnergy}, we may write the rate of change of the semi-major axis in terms of the rate of change of orbital energy
 \begin{equation}
 \frac{d a}{dt} = \frac{G M_\star \mDM}{E^2}\frac{dE}{dt} = \frac{2 a^2}{G M_\star \mDM} \frac{dE}{dt}\ ,
 \label{eq:dadtElliptic}
 \end{equation}
 wherein we can make the same replacements as in the 1-dimensional case, namely that $dE \sim \langle \Delta E \rangle$ as given by Eq.~\eqref{eq:avgdE}, and now using the orbital period for the elliptic orbit in terms of $a$ given above. The quantity $E_{\rm exit}$ in Eq.~\eqref{eq:avgdE}, which we recall is the energy of the DM upon it exiting the star after a stellar transit, is obtained from energy conservation in an elliptic orbit as
 \begin{equation}
 E_{\rm exit} = G M_\star \mDM \left(\frac{1}{R_\star} - \frac{1}{2a}  \right) \ ,
 \end{equation}
 so that we solve Eq.~\eqref{eq:dadtElliptic} to find
 \begin{equation}
 t_{\rm therm, 1} \simeq \frac{\pi}{\sqrt{2}v_{\rm esc,\star} \bar{\rho}_\star \sigma_c} 
\begin{cases}
 \mDM \left( \frac{1}{\sqrt{(2-\epsilon)\epsilon}} - 1 + \frac{\text{artanh}(\sqrt{2}) - \text{artanh}(\frac{\sqrt{2}}{\sqrt{(2-\epsilon)\epsilon}})}{\sqrt{2}} \right)\ , ~ \mDM v_{\rm esc,\star}^2 \gg m_i v_i^2 \\
 m_i \frac{v_{\rm esc,\star}^2}{v_i^2} \left(\frac{1}{\sqrt{(2-\epsilon)\epsilon}} - 1 \right) \ , ~~~~~~~~~~~~~~~~~~~~~~~~~~~~~~~~~ \mDM v_{\rm esc,\star}^2 \ll m_i v_i^2 
  \end{cases} 
   \ .
    \label{eq:tTherm1_elliptic}
 \end{equation}
 This result is similar to that of the 1-dimensional simplified problem, and most importantly, exhibits the same scaling with $\mDM,~m_i$ in the relevant limits.

 The previous result in the literature differs from the above derivation for elliptic orbits in two ways. The first is that Ref.~\cite{Acevedo:2020gro} defines the orbital period in terms of the periapsis distance, $r_p \equiv a(1-e)$ as $\tau = 2\pi \sqrt{r_p^3/G M_\star}$. Defining the orbital period in terms of the periapsis distance requires rescaling by $(1-e)^{-3/2}$, which was not done in Ref.~\cite{Acevedo:2020gro}. Further, the orbital energy was defined in Ref.~\cite{Acevedo:2020gro} in terms of the periapsis distance as $E = - G M_\star \mDM / r_p$, which also neglects a factor of $(1-e)$. Using the periapsis distance, appropriately scaled by factors of $(1-e)$ can be used to derive the thermalisation time, but would require knowledge of the eccentricity $e$. Since this depends on knowing the trajectory of the DM, it is rather impractical. Further, since the energy of the elliptic orbit, and its decrease, can be entirely characterised by the semi-major axis, considering the eccentricity is not required. 

 The other main difference between our result and those in the literature, Refs.~\cite{Kouvaris:2010jy,Acevedo:2020gro}, is the non-inclusion of the gravitational potential contribution to the change of energy in the DM as it transits the star. The reason for this omission in our analysis is that, assuming the DM follows a roughly symmetric trajectory on its way into and out of the star, this contribution averages to zero over the distance travelled in the star. Since we have used the exit energy of the DM after a completed stellar transit in our calculation above, this simplification seems appropriate.

 Note that in the rest of our analysis, we use Eq.~\eqref{eq:tTherm1_elliptic} as opposed to Eq.~\eqref{eq:tTherm1} due to the fact that DM orbits will have a wide range of possible eccentricities, including the special case described by the latter. We choose $\epsilon = 10^{-4}$ to optimise between the capture rate and the time required for thermalisation to occur. Choosing a smaller $\epsilon$ enhances the capture rate, but also increases the thermalisation time, while a larger $\epsilon$ has the opposite effect.

 \subsection{Second thermalisation time}

 Having computed the time required for DM orbits to shrink sufficiently that they are entirely contained in the star, we must now compute the time required for DM to settle into a thermal core in the centre of the star. The process now only depends on the energy loss rate from collisions with the stellar material, Eq.~\eqref{eq:dEdx}, and the relative collisional velocity, which is $v_{\rm rel} \sim c_s$. The initial energy of the DM is that associated with an orbit of radius $R_\star/2$, while full thermalisation is achieved when the kinetic energy reaches $3T_\star/2$. Therefore the second thermalisation time can be shown to be given by
 \begin{align}
t_{\rm therm,2} \simeq \frac{\mDM \log \left[\frac{2 G \mDM M_\star}{3 R_\star T_\star}\right]}{4 \rho_\star^c \sigma_c c_s} + \mathcal{O}(\sqrt{m_i}) \ , ~~~ \mDM \gg m_i \ , \\
t_{\rm therm,2} \sim \frac{m_i^2 \log\,2}{4 \rho_\star^c \sigma_c c_s \mDM} + \frac{m_i\left(\log\,2 + \frac{G M_\star}{c_s^2 R_\star} \right)}{\rho_\star^c \sigma_c c_s} + \mathcal{O}(\mDM)\ , ~~~ \mDM \ll m_i \ .
 \end{align}
 As we will see, this timescale is roughly the same as that which governs the collapse into a black hole.

 \subsection{Large cross-section and diffusive drift}
 \label{sec:Sinking}

 If the DM-SM cross-section $\sigma_c$ is very large, the DM is stopped in the outer regions of the star, and the above discussions of thermalisation do not apply. In this case, the DM will drift towards the core of the star as a result of balancing the stellar gravitational pull, and friction the DM experiences as it moves past the stellar material
 \begin{equation}
 \frac{G M_\star(r) \mDM}{r^2} = \rho_\star(r) c_s(r) \sigma_c v_\chi(r) \ .
 \label{eq:Drift}
 \end{equation}
 This equation can be solved numerically for $v_\chi(r)$ along the trajectory going from approximately the stellar surface to the core. 

 The time taken to reach the core depends on the drift velocity $v_\chi(r)$ as
 \begin{equation}
 t_{\rm sink} \simeq \int_{r_{\rm th}}^{R_\star} \frac{dr}{v_\chi(r)} \ .
 \label{eq:sinkTime}
 \end{equation}
The endpoint of the integral is not $r=0$, which is singular since the solution to Eq.~\eqref{eq:Drift} above gives $v_\chi(0) =0$. Therefore, we integrate as far as the thermal radius, a quantity that will be explained in detail in the next section. Numerically, we find that 
\begin{align}
t_{\rm sink} \sim 1~\text{yr} \times \left(\frac{25~\text{GeV}}{\mDM} \right) \left( \frac{\sigma_c}{10^{-30}~\text{cm}^2} \right) \ ,
\end{align}
valid for $\sigma_c/\mDM \gtrsim (10^{-35}/25)~(\text{cm}^2/\text{GeV})$.

\subsection{Equilibrating inside the star and evaporation}
\label{sec:equilibration}

Once the captured dark matter has thermalised/sunk towards the core of the star, it settles into a thermally-supported sphere. We may use the virial theorem to compute the size of this sphere. The average energy of the DM is $\langle E \rangle = 3 T_\star / 2$, while the gravitational potential energy at the outer envelope of the thermal sphere is $\langle V \rangle = -4 \pi r_{\rm th}^2 \rho_\star G \mDM$. Since the typical size of the thermal radius is $r_{\rm th} \ll R_\star$, we will estimate it assuming $\rho_\star = \rho_\star^c$, the density in the stellar core. Therefore, we arrive at
\begin{equation}
r_{\rm th} = \frac{3}{2}\left(\frac{T_\star}{\pi G \rho_\star^c \mDM }\right)^{1/2} \simeq 10^7~\text{m} \times \left( \frac{10^4~\text{GeV}}{\mDM}\right)^{1/2} \left(\frac{T_\star}{10^8~\text{K}} \right)^{1/2} \left(\frac{6~\text{g/cm}^3}{\rho_\star^c} \right)^{1/2} \ ,
\label{eq:rTherm}
\end{equation}
which should be compared to the typical radius of a Pop III star, $R_\star \sim 5 R_\odot \sim 3\times 10^9~\text{m}$, assuming the benchmark of $M_\star \sim 110 M_\odot$ we consider here.

The thermal radius enters the numerical calculation of the sinking time above, and will also be important in determining the conditions for black hole formation, as we will see in Section~\ref{sec:Collapse}. However, before such a collapse can occur, it is important to consider the possibility that a fraction of the DM will evaporate from the stellar core due to \emph{gaining} energy through collisions with the stellar material. This effect inhibits the permanent capture of DM with mass below roughly the mass of the stellar targets, which we take to be hydrogen. We will find that this is relevant only for the lowest masses that can be probed by Pop III stars.

In practice, one should compute the fraction of DM particles in a given shell at a depth $r$ in the star that is up-scattered to $v> v_{\rm esc}(r)$. The rate at which a DM particle with velocity $w$ scatters to velocity $v$ has been computed in Ref.~\cite{Gould:1987ju}, and depends on the temperature, density and velocity distribution of the target particles. The rate per shell is roughly~\cite{Ilie:2020iup}
\begin{equation}
R(r) \sim \frac{2}{\sqrt{\pi}} n_i(r) c_s(r) \sigma_c \, \text{exp}[-v_{\rm esc}(r)^2/v^2] \ ,
\end{equation}
such that the total evaporation rate can be defined as 
\begin{equation}
E = \frac{\int_{V_\star} n_\chi(r) R(r)}{\int_{V_\star} n_\chi(r)} \ .
\label{eq:EvapRate}
\end{equation}
Assuming that the DM has settled into a thermally-supported sphere as discussed above, its distribution will be
\begin{equation}
n_\chi(r) \sim n_{\chi}(0)\,\text{exp}[-(r/r_{\rm th})^2] \ ,
\end{equation}
where $r_{\rm th}$ is the thermal radius of Eq.~\eqref{eq:rTherm}. The total evaporation rate is then computed numerically. As shown in Fig.~\ref{fig:EddingtonLimited}, evaporation is efficient for masses below roughly $\mDM \lesssim 10~\text{GeV}$, and cross-sections greater than $\sigma_c \gtrsim 10^{-40}~\text{cm}^2$. For cross-sections smaller than this value, the likelihood of up-scattering to above the escape velocity is low, while for masses greater than $10~\text{GeV}$, the DM is heavy enough that the likelihood of up-scattering is exponentially suppressed. Note that at larger cross-sections, up-scattered DM has a high probability of scattering off the stellar medium again. This can lead to additional up-scattering, but also down-scattering. As such, it is reasonable to expect that there is some reduction in the upper bound of $\mDM$ that is evaporated~\cite{Garani:2021feo,GaraniDiscussion}.

 \section{Collapse into a Black Hole}
 \label{sec:Collapse}

 \begin{figure}[t]
 \centering
 \includegraphics[scale=0.8]{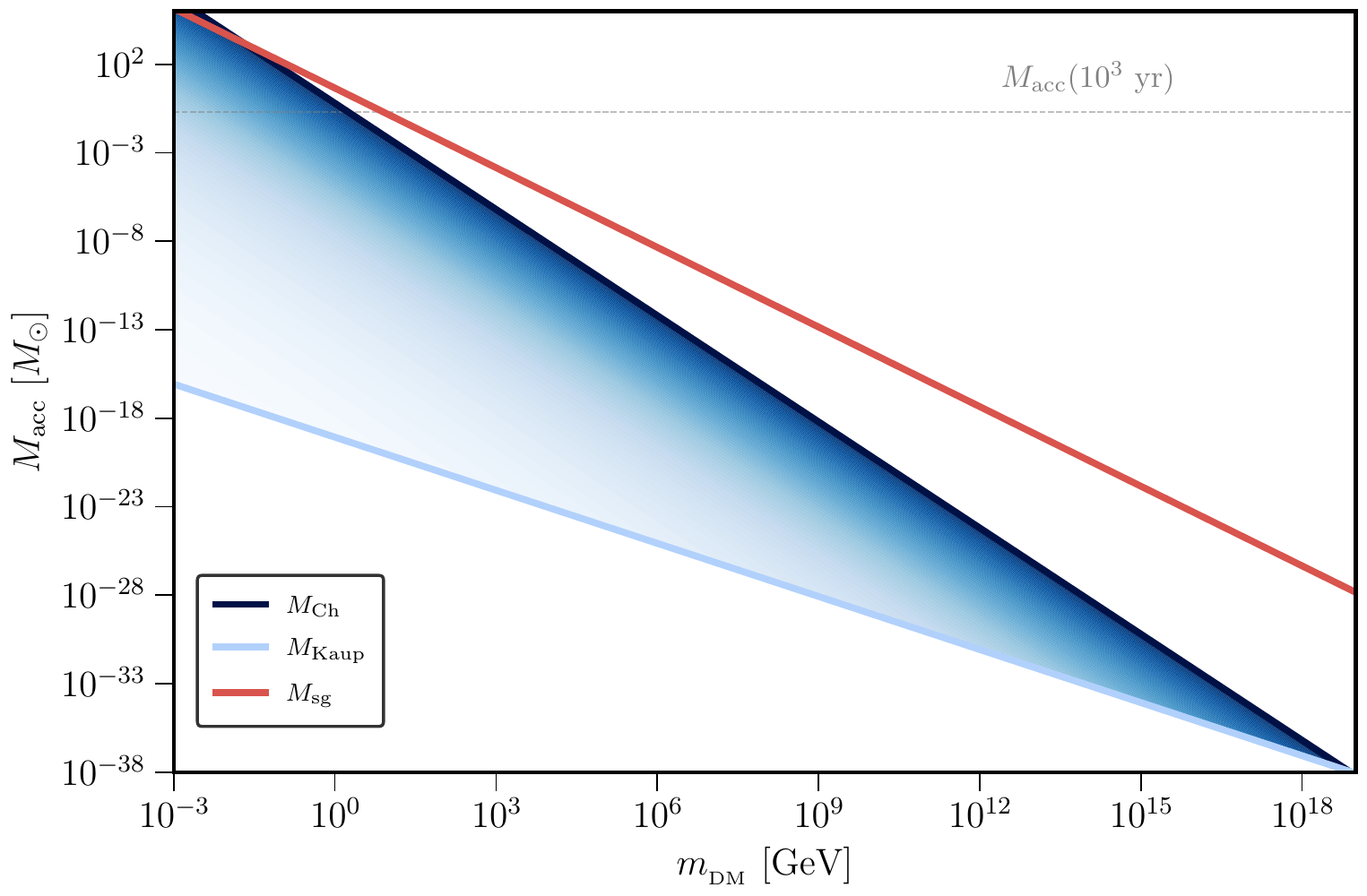}
 \caption{Mass thresholds above which gravitational collapse of the accumulated DM will occur, as a function of the DM mass $\mDM$. If $M_{\rm acc}$ exceeds the largest of these thresholds, DM will collapse into a black hole of mass $M_{\rm BH,0} \sim M_{\rm acc}$. The self-gravitation mass of an isothermal sphere, given by Eq.~\eqref{eq:Msg} is shown in red. The Chandrasekhar mass for fermionic DM is shown in dark blue. Bosonic DM can have a maximal mass bounded from below by the Kaup mass $M_{\rm Kaup} \sim \mP^2/\mDM$ for non-interacting DM, or by $M_{\rm max} \sim \sqrt{\lambda/4\pi} \mP^3/\mDM^2$ for DM with a repulsive quartic self-interaction term $\lambda\varphi^4/4!$~\cite{Chavanis:2011cz,Eby:2015hsq}. We shade the region above $M_{\rm Kaup}$ and below $M_{\rm Ch}\sim M_{\rm max}|_{\lambda = 4\pi}$ to demonstrate our ignorance of $\lambda$.}
 \label{fig:MaxMass}
 \end{figure}

In this section we review the conditions required for collapse of the DM into a black hole, as well as estimate the time required for the collapse to occur. 

 \subsection{Conditions for Collapse}

The thermal sphere is stable as long as $\rho_\chi$, the density of DM in the sphere, is below the density of the surrounding stellar material. When $\rho_\chi \gtrsim \rho_\star^c$, the self-gravitation of the DM will dominate the structure, leading to the gravothermal catastrophe~\cite{Lynden-Bell:1968eqn}. Self-gravitating systems have negative specific heat, so that they lose total energy while heating up and contracting. Since the DM is in contact with a reservoir in the form of the surrounding stellar material, the DM will shed energy, increasing in temperature and collapsing further until eventually it either collapses entirely into a black hole, or the continued collapse is prevented by some other source of pressure. The maximal captured DM mass that can be sustained before this collapse occurs can be found by using the virial theorem as above, modified to include the DM self-gravitation contribution. One finds that a solution for a stable thermal sphere of mass $M_{\rm acc}$ only exists for 
\begin{align}
M_{\rm acc} \lesssim M_{\rm sg} \equiv \sqrt{\frac{3}{\pi \rho_\star^c}}\left(\frac{T_\star^c}{G \mDM}\right)^{3/2} \ .
\label{eq:Msg}
\end{align}

Upon accumulation of $M_{\rm sg}$, formation of a black hole can be prevented if degeneracy pressure begins to dominate over thermal pressure, stabilising the DM against further collapse. This will not occur as long as $M_{\rm sg} \gtrsim M_{\rm Ch}$, the Chandrasekhar mass. For fermionic DM, this quantity is well-known and is~\cite{Kouvaris:2015rea,Gresham:2018rqo}
\begin{align}
M_{\rm Ch} \sim \frac{\mP^3}{\mDM^2} \ ,
\label{eq:Mch}
\end{align}
where we have omitted a prefactor that depends on possible self-interactions.\footnote{Note that per Ref.~\cite{Gresham:2018rqo}, if the DM has self-interactions, the scaling of the maximal mass depends on the mass of the mediator $m_X$ and the coupling strength $g_X$, such that $M_{\rm Ch} \sim \frac{g_X \mP^3}{\mDM m_X}$ up to a numerical coefficient.} 
For bosonic DM formed due to repulsive quartic self-interactions of the form $\lambda \varphi^4/4!$, a similar maximum mass exists, $M_{\rm max} \simeq \sqrt{\lambda/4\pi} M_{\rm Ch}$~\cite{Chavanis:2011cz,Eby:2015hsq}. However, for bosonic DM with no self-interactions, the limiting mass is the Kaup limit~\cite{Kaup:1968zz}
\begin{align}
M_{\rm Kaup} \sim 0.633 \frac{\mP^2}{\mDM} \ .
\label{eq:Mkaup}
\end{align}
These mass scales are shown as a function of the DM mass in Fig.~\ref{fig:MaxMass}.

As long as the captured DM accumulates into a sphere of mass equal to the largest of the limiting mass scales, it can collapse into a black hole unimpeded. In what follows, we will require that $M_{\rm acc} = \text{Max}[M_{\rm sg}, M_{\rm Ch}]$ for black hole formation to begin, since this automatically satisfies the Kaup limit for all choices of $\mDM$.

\subsection{Collapse Time}

Once enough DM is accumulated at the core of the star, collapse begins. We estimate the collapse time according to whether the maximal threshold is the self-gravitating mass or the Chandrasekhar mass. 

If the limiting factor for collapse is the Chandrasekhar mass, the collapse time is on the order of the gravitational free-fall time~\cite{Kippenhahn:1994wva} and almost instantaneous,

\begin{equation}
t_{\rm ff} = \sqrt{\frac{3\pi}{32 G \rho_\chi}} \simeq 10^{-11}~\text{yr} \times \left( \frac{10^{14}~\text{g/cm}^3}{\rho_\chi}\right)^{1/2} \ .
\end{equation}
In this estimate, we have used as a benchmark $\rho_\chi \gg \rho_\star^c$. This is because a degeneracy pressure-supported object at the Chandrasekhar limit will have a density on the order of $\rho_\chi \sim \mDM^4 \sim 10^{14}~\text{g/cm}^3\times(\mDM/1~\text{GeV})^4$.\footnote{In Ref.~\cite{Gresham:2018rqo}, the Tolman-Oppenheimer-Volkov equations were solved for asymmetric DM with various interaction types to find the maximal mass and radius of such objects. Their analysis confirms the expected density and scaling we show here, and matches our expectation from neutron stars, whose core densities are also $\rho_{_{\rm NS}} \sim 10^{14}~\text{g/cm}^3$.} From Fig.~\ref{fig:MaxMass} we see that this timescale for collapse to a black hole is only relevant for DM masses $\mDM \lesssim 10^{-2}~\text{GeV}$, where $M_{\rm Ch} \gtrsim M_{\rm sg}$.

For most of the parameter space we consider, the self-gravitating mass is the limiting parameter. In this case, the time for collapse will be dominated by the requirement that the DM shed its gravitational energy in a timely fashion, as the thermal sphere heats up and shrinks. The initial radius of the sphere is $r_{\rm th}$ above, while the final radius is $r_s$, the Schwarzschild radius. This shedding of energy proceeds through scattering of the DM off the surrounding stellar matter. Thus, the rate of change of energy is given by
\begin{equation}
\frac{dE}{dt} \simeq v_{\rm rel}\frac{dE}{dx} \ ,
\end{equation}
 where $v_{\rm rel}$ is the relative velocity of the DM and the stellar material. Initially the temperatures of the DM and the stellar material are equal, $T_\chi = T_\star^c$. Therefore, for $\mDM \gtrsim\text{GeV}$, $v_{\rm rel} \sim c_s$, the local sound speed, while for $\mDM \lesssim\text{GeV}$, $v_{\rm rel} \sim \sqrt{G M_{\rm sg}/2r}$. Using Eq.~\eqref{eq:dEdx}, and using $\langle E\rangle = \frac{G M_\star}{2r}$ to compute the endpoints of integration as a function of the endpoint radii, we compute the time required for the collapse to be
\begin{align}
t_{\rm coll} \simeq \frac{\mDM \log \left[c_s^2 \frac{r_{\rm th}}{G M_{\rm sg}}\right]}{4 \rho_\star^c \sigma_c c_s} + \mathcal{O}(\sqrt{m_i}) \ , ~~~ \mDM \gg m_i \ , \\
t_{\rm coll} \simeq \frac{m_i \left(\frac{\sqrt{2}}{c_s} - \frac{2}{c_s} \sqrt{\frac{G M_{\rm sg}}{r_{\rm th}}} + \log \left[\frac{r_{\rm th}}{ 2 G M_{\rm sg}}\right]\right)}{4 \rho_\star^c \sigma_c c_s} + \mathcal{O}(\mDM) \ , ~~~ \mDM \ll m_i \ .
\end{align}

For very large cross-sections, collapse can be impeded by drag from the stellar material. This is analogous to the discussion regarding the sinking time in Section~\ref{sec:Sinking}. The calculation of the collapse time proceeds in the same manner, with the sinking velocity $v_\chi(r)$ given by the solution to Eq.~\eqref{eq:Drift} with $M_{\rm sg}$ replacing $M_\star(r)$. The collapse time is then
\begin{align}
t_{\rm coll} \simeq \int_{r_s}^{r_{\rm th}} \frac{dr}{v_\chi(r)} = \int_{r_s}^{r_{\rm th}} dr \frac{\rho_\star(r) c_s(r) \sigma_c r^2}{G M_{\rm sg} \mDM} \ ,
\end{align}
which we solve for numerically. For DM that has collected in a small thermal radius, $\rho_\star(r)$ and $c_s(r)$ are roughly constant across the region of integration, such that the collapse time in the drift regime is approximately 
\begin{align}
t_{\rm coll} \sim \frac{\rho_\star(r_{\rm th}) c_s(r_{\rm th}) \sigma_c (r_{\rm th}^3 - r_s^3))}{3G M_{\rm sg} \mDM} \ .
\end{align}

\section{Black Hole evolution}
\label{sec:BHEvol}

In the previous sections we determined that Pop III stars accumulate large quantities of DM that will self-gravitate and collapse to form a black hole relatively rapidly. Thus, we are now in a position to evaluate the subsequent evolution of the nascent black hole. 

The evolution of the newly-formed black hole is dictated by three rates: the accretion rate due to continued capture of dark matter that reaches the stellar core, the accretion rate of stellar matter onto the black hole, and the Hawking radiation rate at which the black hole loses mass.
\begin{align}
\dot{M}_{\rm BH} = \dot{M}_{\rm cap} + \dot{M}_{\rm acc} - \dot{M}_{\rm H} \ .
\label{eq:BHevol}
\end{align}
The continued capture accretion rate $\dot{M}_{\rm cap}$ is fixed by Eq.~\eqref{eq:CaptureRate}. The rate at which a non-charged non-rotating black hole radiates its mass away is well-known, and can be approximated as
\begin{align}
\dot{M}_{\rm H} \sim \frac{\pi^2}{60} \frac{g_{\rm eff}(T_{\rm H})}{2} \, T_{\rm H}^4 \, (4\pi r_s^2) \ .
\label{eq:Hawking}
\end{align}
Hawking radiation is emitted with an approximately blackbody spectrum, and is therefore characterised by a temperature $T_{\rm H} = (8\pi G M_{\rm BH})^{-1}$.\footnote{For a Kerr black hole, this is modified to $T_{\rm H} = \frac{\sqrt{M_{\rm BH}^2-a^2}}{4\pi G M_{\rm BH}(M_{\rm BH} + \sqrt{M_{\rm BH}^2-a^2})}$, where $a = J/M_{\rm BH}$ is the Kerr parameter given angular momentum $J$. For a Kerr-Newman black hole of electric charge $Q$ and magnetic charge $P$, replace $a^2 \to a^2 + Q^2 + P^2$. We will not consider rotating or charged black holes further here, although it is an interesting possible extension of this work.} The black hole can only radiate particles whose masses are below this temperature, and this factor enters as the effective number of degrees of freedom $g_{\rm eff}(T_{\rm H})$. Finally, the radiation is approximately isotropic from the black hole surface of size $4\pi r_s^2$. 

Recall that the nascent black hole will have a mass given by $M_{\rm sg}$ for $\mDM \gtrsim 10^{-2}$, and therefore scales as $M_{\rm BH} \propto \mDM^{-3/2}$. The Hawking radiation rate scales as $\dot{M}_{\rm BH} \propto M_{\rm BH}^{-2}$, so that at the time of formation, the radiation rate scales as $\mDM^{3}$. At large DM masses, the accretion rate is either independent of $\mDM$ (large $\sigma_c$ where $v_{\rm max} \gtrsim \Delta v_+$), or $\dot{M}_{\rm acc} \propto \mDM^{-1}$ ($v_{\rm max} \lesssim \Delta v_-$). Therefore, for some large value of $\mDM$, the accretion rate will drop below the radiation rate of the nascent black hole. 

For $v_{\rm max} \gtrsim \Delta v_+$, we find the condition for radiation to dominate to be
\begin{align}
\mDM &\gtrsim \mP 
\times 
\left( \frac{4\,\text{km/s}}{v_0} \right)^{1/3} \left( \frac{50}{g_{\rm eff}(T_{\rm BH})} \right)^{1/3} \left( \frac{\rhoDM}{5000\,\text{GeV/cm}^3} \right)^{1/3} \ ,
\end{align}
assuming $M_\star = 110~M_\odot$, $R_\star = 5~R_\odot$, $T_c \sim 10^8~\text{K}$, and $\rho_\star^c \sim 6~\text{g/cm}^3$. The threshold depends on these parameters as $\mDM \propto  T_c (M_\star R_\star /\rho_\star^c)^{1/3}$. Since the DM density in the vicinity of a Pop III star is thought to be as high as $\rhoDM \sim 10^{11}~\text{GeV/cm}^3$, radiation will only dominate over DM accretion for far super-Planckian $\mDM$.

Meanwhile, for $v_{\rm max} \lesssim \Delta v_-$, we find the radiation-domination threshold to be
\begin{align}
\mDM \gtrsim \mP \times \left(\frac{\sigma_c}{10^{-25}~\text{cm}^2}\right)^{1/4} \left( \frac{4\,\text{km/s}}{v_0} \right)^{3/4} \left( \frac{50}{g_{\rm eff}(T_{\rm BH})} \right)^{1/4} \left( \frac{\rhoDM}{5\times10^7\,\text{GeV/cm}^3} \right)^{1/4} \ ,
\end{align}
where we assume the same parameters as above for the star. Unlike the large $v_{\rm max}$ result above, here the threshold scales as $\mDM \propto (G M_\star^3 T_c^3 /R_\star^2 \rho_\star^c )^{1/4}$. Ultimately, despite this scaling, Hawking radiation will not dominate over DM accretion in any of the parameter space we consider.

\subsection{The accretion of stellar matter}

The accretion of the stellar material onto the black hole requires more careful analysis. If the Pop III star was non-rotating, and likewise the black hole at its core, then the accretion rate could be straightforwardly assumed to be given by the Bondi rate~\cite{1952MNRAS.112..195B}
\begin{align}
\dot{M}_{\rm Bondi} = 4\pi \lambda \frac{G^2 M_\star^2}{c_s^3} \rho_\star^c \ ,
\label{eq:BondiRate}
\end{align}
where $\lambda$ is an order one constant. However, it is expected that Pop III stars can be formed with substantial angular momentum~\cite{2013MNRAS.431.1470S}, which could therefore affect this accretion~\cite{Markovic:1994bu,Kouvaris:2013kra}. 

\subsubsection{Impact of rotation on accretion}

If fluid flowing from large radii towards the black hole has large angular momentum, it can be held up at the Kepler radius corresponding to that angular momentum before reaching the black hole. Thus, if the last stable orbit radius is smaller than the Kepler radius of the inflowing fluid, a toroidal accretion structure can form, supported by angular momentum. If this structure has a sufficiently large luminosity, the radiation pressure can hold up material inflow onto the black hole, slowing down the rate of accretion. An extreme example of this would be if the accreting matter reached the Eddington luminosity, which would dramatically slow the accretion rate. However, if there are efficient mechanisms for reducing the initial angular momentum of the inflowing fluid, such as viscous or magnetic braking, it can prevent the formation of a torus with the associated increased luminosity. Our knowledge of the internal magnetic field structure of Pop III stars being limited, we will only consider viscous braking in any detail.

A fluid element at a distance $r$ from the black hole has specific angular momentum $\ell = \omega_\star r^2$. Meanwhile the specific angular momentum of the gas at the innermost stable circular orbit (ISCO) of a black hole is
\begin{align}
\ell_{\rm ISCO} = 2 \sqrt{3}\, \psi\, G M_{\rm BH} \ ,
\label{eq:ISCO}
\end{align}
where $\psi =1,~1/3$ for a Schwarzschild or extreme Kerr black hole respectively. If there exists some $\ell \gtrsim \ell_{\rm ISCO}$, it will stall rather than continue to accrete, contributing to the formation of an accretion torus. Solving for the critical black hole mass $M_{\rm sph}$ above which no such $\ell$ exists, and accretion will be spherically symmetric, we find
\begin{align}
M_{\rm sph} = \frac{\sqrt{3} \, \w_\star^3}{128 \pi^2G^3\,\rho^2\,\psi^3 }  \sim 11 \, M_\odot \times \left(\frac{5 R_\odot}{R_\star} \right)^3\left(\frac{v_{\rm rot}}{400~\text{km/s}} \right)^3 \left( \frac{6~\text{g/cm}^3}{\rho} \right)^2 \left( \frac{1}{\psi} \right)^3\ ,
\label{eq:Msph}
\end{align}
where we have used $\w_\star = v_{\rm rot} / (2\pi R_\star)$. In our estimate we have taken a rotational velocity $v_{\rm rot} \sim 400~\text{km/s}$ as a fiducial value, corresponding to $v_{\rm rot} \sim 0.2~v_{\rm K}$, where $v_{\rm K}$ is the Keplerian velocity associated with a test mass orbiting at the surface of the star. An $M_\star = 110 M_\odot$ Pop III star with $v_{\rm rot} = 0.2~v_{\rm K}$ at ZAMS would be expected to end its life as a PISN~\cite{2012A&A...542A.113Y}, not leaving a black hole remnant. In principle, $v_{\rm rot} \lesssim 0.75~v_{\rm K}$ is allowed for such a star~\cite{2012A&A...542A.113Y}. At this maximal value of $v_{\rm rot}$, $M_{\rm sph} \sim 600 M_\odot$, meaning that an accretion torus would almost certainly exist.

We conclude that in much of the $v_{\rm rot}$ parameter space, an accretion torus is likely to form for at least part of the time during which the accretion occurs, unless there is a mechanism to reduce the angular momentum of infalling material. The viscosity of the stellar material provides such a mechanism, as we will see below.

\subsubsection{Viscous braking}

We follow the discussion of Ref.~\cite{Markovic:1994bu} considering accretion onto a black hole at the centre of our Sun, modified so as to be applied to a Pop III star. Consider a region near the equator of the star of angular size $\alpha$ and thickness $\delta r$, such that its total angular momentum is $\Lagr = \alpha 2\pi \delta r \rho_\star(r) r^2 \ell$. The rate of change of total angular momentum is
\begin{align}
\frac{d\Lagr}{dt} = (\partial_t + v_r \partial_r) \Lagr \simeq - \delta r \partial_r \mathcal{T}(r) \ ,
\end{align}
where $v_r$ is the radial velocity of an infalling fluid element, and $\mathcal{T}(r) = - \alpha 2\pi \rho_\star(r) r^4 \nu (\partial_r \w_\star)$ is the viscous torque in a material of kinematic viscosity $\nu$. The radial velocity can be assumed to be that of a fluid element being accreted at the Bondi rate,
\begin{align}
v_r = - \frac{4\pi \lambda G^2 \rho_\star^c}{c_s^3}\frac{M_{\rm BH}^2}{4\pi \rho_\star(r) r^2 }  = - B \frac{M_{\rm BH}^2}{4\pi \rho_\star(r) r^2 }\ .
\end{align}
Since mass is conserved, we may write
\begin{align}
\partial_t \ell - B \frac{M_{\rm BH}^2}{4\pi \rho_\star(r) r^2 } \partial_r \ell = \frac{1}{\rho_\star(r) r^2} \partial_r \left( \rho_\star(r) r^4 \nu  \partial_r \left(\frac{\ell}{r^2} \right)\right) \ .
\end{align}
Treating the flow as approximately stationary, we arrive at
\begin{align}
\ell + \frac{4\pi \rho_\star(r) \nu}{B M_{\rm BH}^2}r^4 \partial_r \left(\frac{\ell}{r^2}\right) = k \ ,
\end{align}
where $k$ is a constant of integration. For viscous braking to be effective, the second term must dominate, requiring
\begin{align}
r \gtrsim r_\nu = \frac{B M_{\rm BH}^2}{4\pi \rho_\star(r) \nu} \ .
\end{align}
The kinematic viscosity of the stellar material is roughly $\nu \simeq \lambda c_s \simeq 5\times 10^4 ~ \text{cm}^2/\text{s} \times \left(\frac{T}{10^8~\text{K}} \right)^{5/2} \left( \frac{6 \text{g/cm}^3}{\rho_\star} \right)$, where $\lambda \propto (T^2/\rho_\star)$ is the mean free path of a proton experiencing Coulomb deflections. Using the above $r_\nu$ in the expression for $\ell$, and equating it to $\ell_{\rm ISCO}$, we find that viscous braking is efficient in our benchmark Pop III star for
\begin{align}
M_{\rm BH} \lesssim M_\nu  = \frac{c_s^2}{G} \left( \frac{2\sqrt{3}\nu^2 \psi}{\w_\star}\right)^{1/3} \sim 4\times10^{-9}~M_\odot \times \left( \frac{400~\text{km/s}}{v_{\rm rot}} \right)^{1/3} \ .
\label{eq:Mvisc}
\end{align}

Thus, for $M_\nu \lesssim M_{\rm BH} \lesssim M_{\rm sph}$, the accretion rate might be expected to deviate from the Bondi rate, given that the stellar viscosity is insufficient to transfer angular momentum outwards.

Besides viscous braking, there can also be magnetic braking. This mechanism functions by virtue of magnetic fields resisting flow that deforms field lines. Requiring that the magnetic torque per unit area exceed the angular momentum current density would ensure efficient magnetic braking. This would place a condition on the size and distribution of the magnetic fields in the star. Without good knowledge of the magnetic fields in Pop III stars, we do not estimate the efficiency of magnetic braking here.

\subsubsection{Viscous braking in Dark Matter}

Above, we considered the impact of viscous braking on efficient accretion of stellar matter onto the black hole. Through its interaction with stellar matter, dark matter may also end up with too much angular momentum and no way to shed it, leading to slower-than-Bondi accretion. We can estimate the black hole mass for which viscous braking of dark matter accretion is efficient in the same way as above for the stellar matter. The important difference is that the viscosity of DM is
\begin{align}
\nu_{\rm DM} = \lambda_{\rm DM} v_{\rm rel}  = \frac{v_{\rm rel}}{n_i \sigma_c}\ ,
\end{align}
where $\lambda_{\rm DM}$ is the mean free path of DM in the stellar fluid, and therefore depends on the number density of target particles $n_i$ and the interaction cross-section $\sigma_c$. Meanwhile $v_{\rm rel} \sim |v - c_s|$ is the relative velocity in the collisions between the two species. For large DM masses this will be $c_s$, while for small masses it will be $v \sim \sqrt{T/\mDM}$.

To compute $M_\nu^{\rm DM}$, we must know the velocity of DM in the direction around the axis of the black hole's rotation. We assume the black hole co-rotates with the star. Since we are assuming that when the DM is falling into the black hole, it has already fully thermalised with the star, it is moving around randomly with $v_{\rm rot} \sim \sqrt{T/\mDM}$. Using the appropriate quantities for our benchmark Pop III star, we find that
\begin{align}
M_\nu^{\rm DM} \sim 110 M_\odot \times 
%\left(\frac{\mDM}{5\times 10^5~\text{GeV}} \right)^{2/3} 
\left(\frac{10^{-36}~\text{cm}^2}{\sigma_c} \right)^{2/3} \left( \frac{\mDM}{10^3~\text{GeV}} \right)^{1/6}\ ,
\label{eq:MviscDM}
\end{align}
for $\mDM \gg \text{GeV}$.

At first glance, this dependence on the cross-section might seem counter-intuitive. However, it is known from Chapman-Enskog theory that the viscosity of gases \emph{increases} with the mean free path~\cite{1970mtnu.book.....C}. This can be reconciled with our intuition by considering the limit where the DM has no interaction with the stellar matter at all. Infalling DM with small initial angular momentum satisfying $\ell \leq \ell_{\rm ISCO}$ will not be held up at the innermost stable radius before being accreted onto the black hole, as it cannot gain additional angular momentum through interactions with the stellar material. Naturally, it cannot gain angular momentum from gravity, a central force, either. Dark matter with self-interactions could change this picture. However, this introduces model-dependence, so we do not consider this possibility further here.

In analogy to the analysis for the stellar matter above, there is also a mass of the black hole above which there is no specific angular momentum of the accreting DM which satisfies $\ell \geq \ell_{\rm ISCO}$. It is given by Eq.~\eqref{eq:Msph}, only with $v_{\rm rot}\sim \sqrt{T/\mDM}$, since the DM is assumed to have already thermalised when it is accreted onto the black hole. Therefore we have
\begin{align}
M_{\rm sph}^{\rm DM} = 66M_\odot \left(\frac{1~\text{GeV}}{\mDM}\right)^{3/2} \ .
\label{eq:MsphDM}
\end{align}
Note that per the scaling of Eqs.~\eqref{eq:MviscDM} and \eqref{eq:MsphDM}, depending on $\sigma_c$, one can be in a situation where $M_{\rm sph}^{\rm DM} < M_{\nu}^{\rm DM}$. In this case, accretion of DM will always remain spherical. The condition for this to occur is that
\begin{align}
\mDM \gtrsim 400~\text{GeV} \left(\frac{\sigma_c}{10^{-30}~\text{cm}^2}\right)^{2/5} \ .
\end{align}

\subsubsection{Potential deviation from Bondi accretion}

For Bondi accretion onto a black hole in a material with large optical thickness, it was shown by Flammang in Ref.~\cite{1984MNRAS.206..589F} that the luminosity due to the infalling matter is
\begin{align}
L_{\rm Flammang} = 4 \left( \frac{\gamma-1}{\gamma}\right)\frac{P_{\rm rad}}{P_{\rm gas}} \frac{4\pi G M_{\rm BH}}{\kappa} \ ,
\end{align}
where $\gamma$ is the adiabatic constant, $P_{\rm rad} \simeq (4/3)\sigma T^4 $ is the pressure due to radiation, and $P_{\rm gas}$ is the total gas pressure. The radiative opacity of the material $\kappa$ is taken to be the Thomson opacity, $\kappa = 0.2(1+X)~\text{cm}^2/\text{g} \sim 0.35~\text{cm}^2/\text{g}$ assuming a Hydrogen fraction $X=0.75$.

Meanwhile, it is well known that there is an upper limit to the luminosity of a system, the Eddington luminosity. For accretion onto a black hole, this limit enforces balance between infalling matter and radiation pressure due to the infall, thereby leading to a dramatically slower accretion rate. The Eddington luminosity is
\begin{align}
L_{\rm Edd} = \frac{4\pi G M_{\rm BH}}{\kappa} \ .
\end{align}
We see that the Flammang luminosity is only smaller than the Eddington luminosity by a factor $4(1-1/\gamma)(P_{\rm rad}/P_{\rm gas}) \sim 0.9$ for a Pop III star. This should be compared with a factor $\sim 10^{-4}$ for a solar-type star. If the Eddington limit is reached while a fraction $\epsilon$ of the energy of the accreted matter is radiated away, we can infer an upper limit on the accretion rate of
\begin{align}
\dot{M}_{\rm BH,~Edd} = \frac{4\pi G M_{\rm BH}}{\epsilon\kappa} \ ,
\label{eq:EddAccretion}
\end{align}
where $\epsilon$ depends on the dynamics of the black hole. For a Schwarzschild black hole, $\epsilon \sim 0.06$, while for a Kerr black hole, $\epsilon \lesssim 0.4$~\cite{1973A&A....24..337S}. Irrespsective of the precise value of $\epsilon$, since the accretion time for Eddington-limited black hole growth is logarithmic in $M_{\rm BH}$, as opposed to the $M_{\rm BH}^{-1}$ dependence of Bondi accretion, this can have a significant impact on the evolution of the black hole. 

Super-Eddington luminosities can be achieved, for example if the accretion disc is geometrically thick and optically thin (a ``Polish Doughnut'')~\cite{Paczynski:1979rz}. However, such models typically predict only linear enhancements of the accretion rate (see e.g. Ref.~\cite{2007MNRAS.377.1187P}). A geometrically thin accretion disk, on the other hand, seems unlikely to form in the stellar interior without being accompanied by complete disruption of the star. More likely, a rather tall toroidal structure could form when viscous braking is no longer effective. Radial accretion would still be supported at the poles, since $\ell \propto \sin\theta^2$, such that for some value of the polar angle $\theta$, $\ell < \ell_{\rm ISCO}$.

Finally, it should be noted that in recent numerical simulations of accretion onto a black hole at the core of neutron stars, it was found that the accretion rate tends not to deviate from Bondi~\cite{East:2019dxt}. However, the interior and consequently the hydrodynamics of neutron stars differs significantly from that of a Pop III star. Two differences in particular are important: neutron stars can have very large magnetic fields, which could contribute to magnetic braking, preserving Bondi accretion; neutron stars can be spinning very rapidly, reducing the impact of viscous braking, leading to deviation from Bondi accretion. Therefore, while this study offers hope that Bondi accretion may be a good approximation for some stellar systems, we cannot conclude with any certainty that it implies the same for Pop III stars. 

We conservatively take the accretion rate between $M_\nu$, given in Eq.~\eqref{eq:Mvisc} and $M_{\rm sph}$ (Eq.~\eqref{eq:Msph}) to be given by the Eddington-limited accretion rate of Eq.~\eqref{eq:EddAccretion}. For dark matter, $M_{\nu}^{\rm DM}$ of Eq.~\eqref{eq:MviscDM} is cross-section and mass-dependent, while $M_{\rm sph}^{\rm DM}$ is mass-dependent. If $M_{\rm sph}^{\rm DM} < M_{\nu}^{\rm DM}$, we assume accretion is governed by the usual rate of Eq.~\eqref{eq:CaptureRate}, while if $M_{\rm sph}^{\rm DM} > M_{\nu}^{\rm DM}$, we assume accretion is Eddington-limited between these two mass scales, where in Eq.~\eqref{eq:EddAccretion} we use $\kappa = \sigma_c/m_i$. 
In practice the lower limit in mass above which Eddington accretion occurs is $\text{Max}[M_{\rm sg},M_\nu^{\rm DM}]$. Therefore, Eddington-limited accretion of DM ultimately only occurs in a small sliver of the parameter space.

\subsection{Summary of Black Hole Evolution}
\label{sec:BHEvolSummary}

\begin{figure}[t]
\centering
\includegraphics[scale=1]{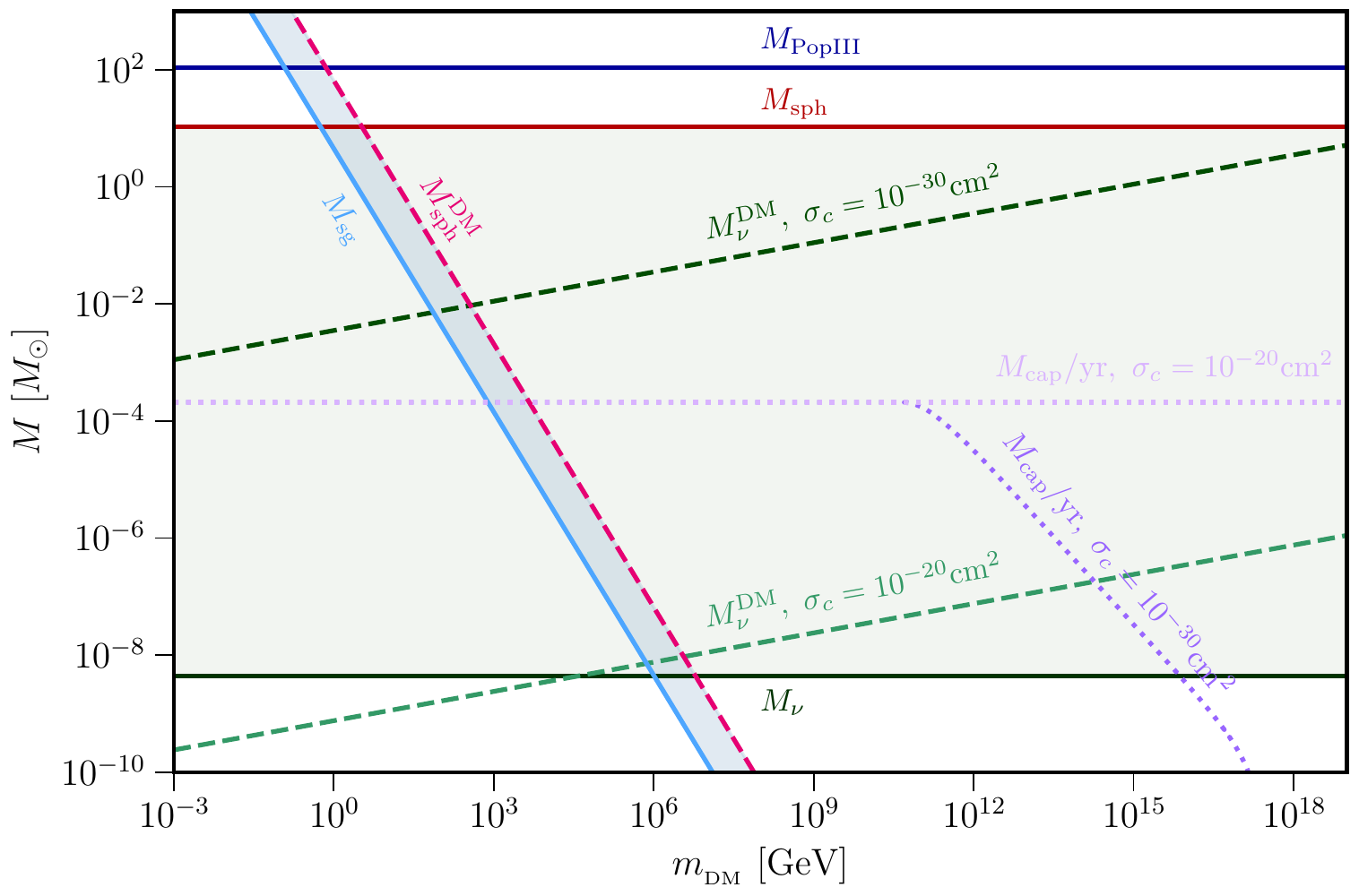}
\caption{The various mass scales in the accretion problem, labelled according to their colour. The benchmark Pop III star mass is $M_{\rm Pop III} = 110\,M_\odot$ and is shown in dark blue. The spherical accretion mass of the stellar material $M_{\rm sph}$ is defined in Eq.~\eqref{eq:Msph}, and is shown in dark red. The spherical accretion mass of dark matter $M_{\rm sph}^{\rm DM}$ is defined in Eq.~\eqref{eq:MsphDM}, and is shown in dashed red. The viscous braking mass for stellar matter $M_\nu$ is defined in Eq.~\eqref{eq:Mvisc} and is shown in dark green. The viscous braking mass for dark matter $M_\nu^{\rm DM}$ is defined in Eq.~\eqref{eq:MviscDM} and is shown in dashed dark (light) green for $\sigma_c = 10^{-30}~(10^{-20})~\text{cm}^2$. The self-gravitation mass of dark matter that collapses to form a black hole, $M_{\rm sg}$ defined in Eq.~\eqref{eq:Msg}, is shown in light blue. Finally, the mass of dark matter accreted (Eq.~\eqref{eq:CaptureRate}) in a year is shown in dotted dark (light) purple for $\sigma_c = 10^{-30}~(10^{-20})~\text{cm}^2$. As explained in Section~\ref{sec:BHEvol}, between $M_\nu^{\rm (DM)}$ and $M_{\rm sph}^{\rm (DM)}$, (dark) matter can be accreted non-spherically. This is shown by the shaded green region for stellar matter, and the shaded blue region for dark matter. Note that for the latter, the lower bound is $\text{Max}[M_{\rm sg},M_{\nu}^{\rm DM}]$ since until the black hole is formed, the accretion rate is unaffected.}
\label{fig:MassScales}
\end{figure}

We consider the evolution of the nascent black holes across mass scales where accretion can be either spherical or Eddington-limited. For ease of reading, we provide here a brief summary of the evolution in both cases, as well as the reasoning behind both limits. The relevant mass scales are shown in Fig.~\ref{fig:MassScales}, and the references to the appropriate equations from the text are in the caption.

The evolution of the black hole can be succinctly summarised as behaving according to Eq.~\eqref{eq:BHevol}. What varies between the cases of spherical/non-spherical accretion are the rates $\dot{M}_{\rm cap}$, dictating the rate at which DM is added to the black hole, and $\dot{M}_{\rm acc}$, the rate at which stellar material is accreted onto the black hole. The Hawking radiation rate, given by Eq.~\eqref{eq:Hawking} is identical in both cases, as it only depends on the temperature (and therefore mass) of the black hole.

\textbf{Spherical Accretion}

Spherical accretion will occur when the infalling matter (dark or stellar) does not have sufficient specific angular momentum to get held up at the innermost stable circular orbit (ISCO) of the black hole, i.e. $\ell \lesssim \ell_{\rm ISCO}$ as given in Eq.~\eqref{eq:ISCO}. Given material orbiting the black hole at some frequency $\w$, this requirement is satisfied for any black hole whose mass exceeds $M_{\rm sph}$ as given in Eq.~\eqref{eq:Msph} for matter, or Eq.~\eqref{eq:MsphDM} for DM. These quantities can be different since the rotational velocities of the stellar matter and DM need not be the same. Spherical accretion can also occur if there is a mechanism for efficiently transferring angular momentum away from the ISCO to outer radii. Viscous braking provides such a mechanism, and is efficient for certain black hole masses. Spherical accretion prevails for $M_{\rm BH} \lesssim M_\nu^{\rm (DM)}$ for stellar (dark) matter, with $M_\nu^{(\text{DM})}$ given in Eq.~\eqref{eq:Mvisc} for stellar (Eq.\eqref{eq:MviscDM} for dark) matter. The key differences between the quantities $M_\nu$ and $M_\nu^{\rm DM}$ lie in the rotational velocities of the respective species, and in the viscosity, which additionally depends on the mean free path of the species. For DM, we have taken the mean free path for collisions with the stellar material as the only source of viscous braking, ignoring potential model-dependent self-interactions.

\textbf{Non-spherical Accretion}

When the black hole mass lies between these two values, $M_\nu^{\rm (DM)} \lesssim M_{\rm BH} \lesssim M_{\rm sph}^{\rm (DM)}$, the dynamics of accretion will depend on the nature and luminosity of the resultant accretion disk. Stellar material is unlikely to form a thin, optically thick disk, as this would constitute a large disruption of the stellar interior. More likely, a toroidal accretion structure would form within the star, and the accretion rate would lie somewhere between the maximal Bondi rate and the minimal Eddington-limited rate. Dark matter, on the other hand, could form a typical accretion disk without disrupting the star. The worst-case scenario is that both stellar and dark matter are accreted at the Eddington-limited rate, Eq.~\eqref{eq:EddAccretion}. 

We will present results conservatively assuming Eddington-limited non-spherical accretion applies, as they end up not being very different from the results if we assumed spherical accretion throughout.

\section{The Fate of Population III Stars}
\label{sec:Fate}

\begin{figure}[t]
\centering
\includegraphics[scale=1]{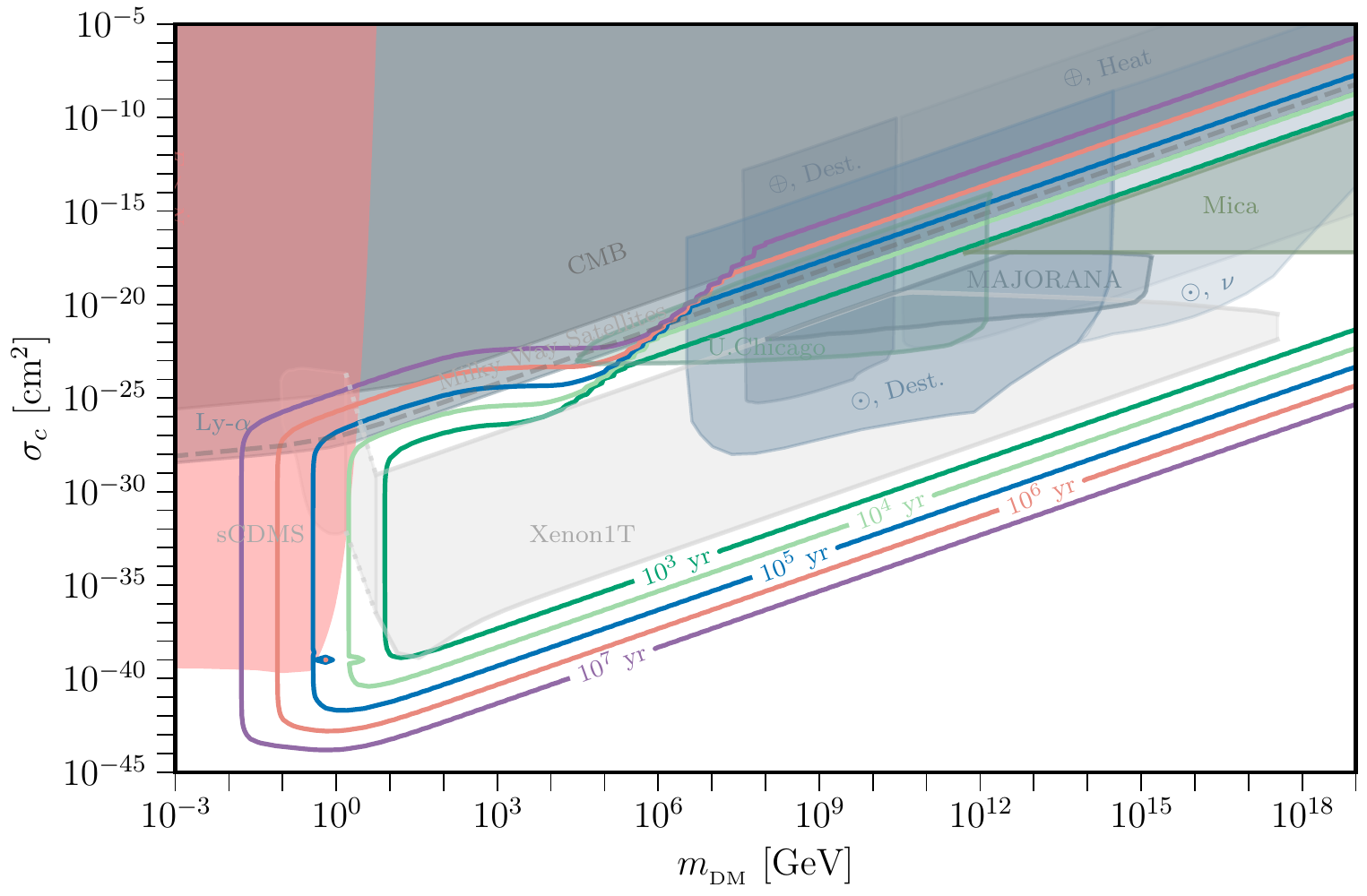}
\caption{Isochrones of time required for enough dark matter to accumulate, collapse into a black hole, and fully consume a $110\,M_\odot$ Pop III star, assuming non-spherical accretion in the appropriate regions of parameter space. For reference, the typical lifetime of a Pop III star of such a mass is $\tau \sim 10^6~\text{yr}$, such that within the red contour, \emph{most} Pop III stars would end their lives as black holes, and not in supernovae as expected in standard stellar evolution theory. The region affected by non-spherical accretion onto the black hole is seen in the top isochrones, for dark matter masses ranging from $100~\text{GeV} \lesssim \mDM \lesssim 10^8~\text{GeV}$, as a function of cross-section $\sigma_c$. See Section~\ref{sec:BHEvol} for details, and a short summary in Section~\ref{sec:BHEvolSummary}. Shaded in light red is the region of parameter space where the evaporation rate of Eq.~\eqref{eq:EvapRate} exceeds the capture rate of dark matter (see Section~\ref{sec:equilibration} for details). In this plot we have taken $\rhoDM \sim 10^{11} \text{GeV/cm}^3$, corresponding to an adiabatically-contracted NFW profile forming at $z\sim 15$~\cite{Ilie:2020iup}. Excluded regions are shown as in Fig.~\ref{fig:velRegime}. Direct observation of a PISN could be used to infer a limit on all the parameter space within an isochrone of $\sim10^5\,\text{yr}$. The parameter space could also be indirectly probed by measurements of the 21-cm signal, or gravitational wave events involving the black hole remnants.}
\label{fig:EddingtonLimited}
\end{figure}

Having discussed how non- or weakly-annihilating dark matter  accumulates in the core of Pop III stars, whereupon it can collapse into a black hole, we now turn to a discussion on the fate of these stars. This depends on the dark matter mass and the interaction cross-section with the material, as we have seen.

In Fig.~\ref{fig:EddingtonLimited}, we see the isochrones of the number of years required for accumulated DM to turn into a black hole and consume the entire Pop III star. The various boundaries of the isochrones can be easily interpreted. The left boundary comes about because at low DM masses, the amount of mass required to collapse into a black hole continually increases, as in Fig.~\ref{fig:MaxMass}. Given the rate of DM capture of Eq.~\eqref{eq:CaptureRate}, eventually the limiting factor becomes the time required to capture the amount of mass that can undergo collapse. The lower boundary arises due to the inefficiency of captured DM to thermalise with the star. It was seen in Section~\ref{sec:Thermalisation} that as the cross-section with matter decreases, the time required for gravitationally captured DM to settle inside of the star increases. Eventually, the time required for this process becomes longer than the natural lifetime of the Pop III star. Finally, the upper boundary has two distinct behaviours -- one that is smooth as a function of $\mDM$, and a kink. Dark matter with large $\sigma_c$ will not only be gravitationally captured, but will come to rest in the outer regions of the star. It must then drift under the competing effects of gravity and drag as seen in Section~\ref{sec:Sinking}, before reaching the core and building up a critical mass to collapse. The smooth part of the upper boundary therefore arises from this sinking time, which can eventually exceed the natural lifetime of the star. The kink is caused by non-spherical accretion of DM onto a nascent black hole. As discussed in Section~\ref{sec:BHEvol}, DM with large cross-sections with matter has low viscosity, and therefore inefficient viscous braking. Therefore, for a certain time between when the nascent black hole reaches $M_\nu^{\rm DM}$ and when it reaches $M_{\rm sph}^{\rm DM}$, we assume Eddington-limited accretion, which slows down the rate at which the star is consumed. 

In the region of Fig.~\ref{fig:EddingtonLimited} shaded in red, it is unlikely that a black hole would form, due to evaporation of captured dark matter. As discussed in Section~\ref{sec:equilibration}, the dark matter continues to collide with the stellar matter after being captured. For masses $\mDM \lesssim 1~\text{GeV}$, the likelihood that dark matter gets upscattered to a velocity above the escape velocity of the star can be signifcant depending on the cross-section $\sigma_c$. We observe a cutoff at $\sigma_c \lesssim 10^{-40}\,\text{cm}^2$ below which evaporation is inefficient due to infrequent scattering.

Finally, we comment on the effect of our choice of benchmark parameters $\rhoDM$, $v_0$, $M_\star$ and $R_\star$. Varying $\rhoDM$ affects the leftmost endpoint of the contours of Fig.~\ref{fig:EddingtonLimited}, with the minimum $\mDM$ constrained scaling as $\mDM^{\rm min} \propto \rhoDM^{-2/3}$ when $v_{\rm max} \gtrsim \Delta v_+$, owing to the scaling of $M_{\rm sg}$ with $\mDM$. At this endpoint, the limiting factor is the requirement that enough DM be captured to reach $M_{\rm sg}$. Meanwhile, changing $M_\star$ and $R_\star$ primarily affect the constrained region through the modified rate of DM accretion, given by Eqs.~\eqref{eq:captureBigvMax}, \eqref{eq:captureSmallvMax} in the relevant incoming DM velocity regimes. Since Pop III stars are thought to have a non-linear scaling of $M_\star$ with $R_\star$, changing these also impacts their core density. Smaller, lighter Pop III stars will tend to have a greater density, decreasing the thermalisation times (Eq.~\eqref{eq:tTherm1}), and $M_{\rm sg}$ (Eq.~\eqref{eq:Msg}). Thus, while smaller and more dense Pop III stars will accrete DM less efficiently, the DM will take less time to thermalise, and not as much needs to be accreted to reach the self-gravitation threshold. There are further, more subtle effects on the accretion of stellar matter, as discussed in Section~\ref{sec:BHEvol}. Finally, our choice of $v_0$ predominantly affects the accretion rate and therefore the low-mass endpoint. In the $v_{\rm max} \gtrsim \Delta v_+$ regime, the scaling of the endpoint goes as $\mDM^{\rm min} \propto v_0^{2/3}$. Meanwhile in the $v_{\rm max} \lesssim \Delta v_-$ regime, $\mDM^{\rm min} \propto v_0^{2}$ if $\mDM v_{\rm esc}^2 \ll m_i c_s^2$, and $\mDM^{\rm min} \propto v_0^{6}$ otherwise. However, the dispersion velocity of the DM in the minihalo cannot vary by more than an order one factor, as the escape velocity for the halo itself is not large compared to $v_0$.

It is apparent from Fig.~\ref{fig:EddingtonLimited} that in a wide range of parameter space, spanning $0.1\,\text{GeV} \lesssim \mDM \lesssim \mP$, and $10^{-42} \lesssim \sigma_c/\text{cm}^2 \lesssim 10^{-8}$, enough dark matter will be captured, settle in the core, collapse into a black hole, and accrete the entire star within $\tau = 10^5~\text{yr}$. This timescale is at least an order of magnitude shorter than the typical lifetime of Pop III stars, and would therefore significantly disrupt the standard expectation of Pop III remnants, which we return to in the next section.

\section{Premature Death of Population III Stars and Reionisation}
\label{sec:DeathImpact}

In the above sections, we have discussed the capture of dark matter onto Pop III stars and their subsequent premature death as black holes. Previously, in Section~\ref{sec:Reionisation} we discussed the global 21-cm signal, and commented on how Pop III star histories could play a role. We now consider the impact of premature black hole death of these early stars on the epoch of reionisation.

Population III stars are efficient UV emitters. This emission couples to the baryon spin temperature through the Wouthuysen-Field effect and the coefficient $x_\alpha$ (see Eq.~\eqref{eq:Tspin}). This coefficient is affected both by the Pop III star formation rate and the duration of time for which Pop III stars are efficiently emitting in the UV. We have not considered the impact of DM on the star formation rate, although this can impact the onset of radiative coupling of the 21-cm transition. The mechanism presented here of premature black hole death would impact the duration of UV emission, which in turn affects both the onset and depth of the global signal. Quantifying this impact requires modelling the UV emission from Pop III stars with truncated lifetimes, an analysis beyond the scope of this paper. In an extreme case, however, one could consider the possibility that the absorption signal might only be very weak due to a dramatic absence of UV radiation during this era. This is inconsistent with the reported EDGES signal~\cite{Bowman:2018yin}, which remains to be confirmed. Confirmation of an absorption signal could be used to place a limit on the truncated stellar lifetime.  

While the impact of premature black hole death on the UV emission can be important, it was argued in Ref.~\cite{Mirocha:2017xxz} that it was the effect of Pop III remnants that could be more visible than the effect of the stars themselves. 
The black hole remnants of Pop III stars can emit both in X-rays and radio as a result of accretion. Emission of X-rays heats the intergalactic medium, which tends to decrease the size of the 21-cm absorption signal by increasing $T_K$. It can also elongate the epoch of reheating, resulting in a gradual transition from an absorption signal to emission, yielding a broad and asymmetric signal~\cite{Mirocha:2017xxz}. Again, this is inconsistent with the reported EDGES result, which await confirmation. The luminosity in X-rays depends on the mass of the black hole, and the fraction of the corresponding radiation that is emitted in the X-ray band. It also depends on the efficiency of accretion onto the black hole, which in binary systems is expected to be roughly $10\%$~\cite{1973A&A....24..337S}.\footnote{The efficiency of accretion onto black holes was discussed in greater detail in Section~\ref{sec:BHEvol}, although the production of X-rays was not commented on there.} The X-ray background can however be suppressed if the halos containing the accreting black holes have column densities exceeding $\text{N}_{\rm H} \sim5\times10^{23}~\text{cm}^{-2}$~\cite{2018ApJ...868...63E}. Such column densities might be achievable, especially if there are no supernovae expelling gas from the halo~\cite{Mebane:2020jwl}.

Radio emission from the Pop III remnants, if sufficiently large, can enhance the size of the signal by raising the background temperature $T_{\rm rad}$. Unfortunately, radio emission from accreting black holes is not well understood, possibly resulting from separate populations of loud and quiet emitters. Ref.~\cite{2018ApJ...868...63E} assumes a population that is $10\%$ radio-loud and $90\%$ radio-quiet, with an empirical luminosity function given in terms of the X-ray luminosity of the accreting black holes. This approach is followed in Ref.~\cite{Mebane:2020jwl}, and it was found in both studies that Pop III star remnants could significantly enhance $T_{\rm rad}$ at redshifts $z\lesssim 20$. However, it should be noted that while enhanced radio emission can match the depth of the observed EDGES signal, the recoupling of the spin and background temperatures occurs via X-ray heating, which is a continuous process. Therefore, while Pop III stars alone can explain the depth and start time of the EDGES signal, they are unlikely to explain the observed sharp ending of the absorption signal.

Another important impact of premature black hole death is the predicted absence of death by supernova. Supernovae, and particularly PISN, release so much energy that they can potentially unbind all the gas in the host halo. This can lead to the expulsion of metals, leading to further Pop III star formation, thereby delaying the start of Pop II formation. Planck results on the total integrated optical depth to reionisation can be applied to constrain this scenario, although it was found in Ref.~\cite{Mirocha:2017xxz} that substantial leeway existed.  Ref.~\cite{Schauer:2019ihk} showed that the timing of the EDGES result could be explained by a Pop III-only model, consistent with significantly delayed Pop II star formation. Finally, it should be noted that the non-occurrence of supernovae in star-forming halos would alter their chemical composition, as none of the heaviest metals could be produced. These alterations of the standard evolutionary history and the prevention of metal formation would not only impact Pop II star formation, but could have knock-on effects on current stellar populations. A detailed study of the propagation of metal non-formation to present times could concievably rule out part of the considered parameter space.

In summary, Population III stars, through their standard evolution, have an important impact on reionisation, and therefore the global 21-cm signal~\cite{1967AZh....44..295D,1977Ap&SS..48..145H,1984ApJ...277..445C,Mirocha:2017xxz,2018ApJ...868...63E,Schauer:2019ihk,Mebane:2020jwl}. Modifying their evolution, as would occur if dark matter were to cause premature black hole death, would significantly impact reionisation. Various possible outcomes are possible, depending on the timing of the premature death, the lack of violent supernovae, and the impact on metal enrichment of the surrounding medium. The details are beyond the scope of this study, but provide a number of avenues for future work.

\section{Consequences of Premature Black Hole Death}
\label{sec:Consequences}

In Sections~\ref{sec:PopIII} and~\ref{sec:Reionisation}, we discussed the salient features of the standard Pop III lifecycle and the global 21-cm signal. In Sections~\ref{sec:Capture},~\ref{sec:Thermalisation},~\ref{sec:Collapse} and~\ref{sec:BHEvol}, we showed how dark matter could be captured by Pop III stars, and result in black holes consuming them from the inside out. In Section~\ref{sec:Fate}, we saw that across a wide range of dark matter masses and interaction cross-sections, the time required to turn the Pop III star into a black hole could be less than the stellar lifetime -- \emph{premature black hole death}. The consequences of this premature death on the 21-cm global signal were discussed in Section~\ref{sec:DeathImpact}. We saw that substantial impact is expected, ranging from signatures that can be excluded when a 21-cm absorption signal is fully confirmed, to knock-on effects that could be constrained by observations of present-day stars.

The consequences of such premature death will in part depend on the required timescale. For example, if Pop III stars are destroyed well before their usual expiry date, they might emit insufficient UV radiation to trigger a measurable absorption feature during reionisation via the Wouthuysen-Field effect. Furthermore, the dearth of stars ending their lives as supernovae would severely impact the transition to the formation of Pop II stars. This scenario is almost certainly ruled out, especially if the EDGES result is confirmed. However, further detailed study needs to be carried out to see if this worst-case scenario can occur as a result of premature black hole death. If, however, the timescale required to consume the star is close to the stellar lifetime, it is likely that fluctuations of the various stellar and DM parameters could play an important role in determining whether all, most, or none of the stars end in premature black hole death. In between these two extremes, premature black hole death would prevent supernova death, leading to a greater number of more massive black hole remnants in host halos. This would lead to additional accretion, resulting in increased X-ray and radio emission. The lack of supernovae would enhance the column density of hydrogen, potentially preventing X-ray heating of the intergalactic medium. This could lead to an enhanced absorption feature~\cite{2018ApJ...868...63E,Mebane:2020jwl}. If the X-rays are not blocked in the host halos, this could lead to a reduced absorption feature due to heating of the gas.

As long as a significant number of Pop III stars experience premature black hole death, there could additionally be implications for experiments searching for gravitational waves. For example, a recently observed merger event is consistent with black holes that cannot have formed as first-generation stellar remnants in the standard evolutionary paradigm~\cite{PhysRevLett.125.101102}. Both new physics~\cite{Sakstein:2020axg} and previous merger history have been suggested as possible explanations. Since premature black hole death could affect Pop III stars of all masses, this would provide a natural mechanism for producing first-generation black holes that could fit the observed event. It remains to be examined whether premature black hole death would result in a merger rate that is consistent with the current rate of observations.

Premature black hole death would also likely preclude the observation of a PISN, a target for the James Webb Space Telescope. If a PISN observation was confirmed, this could provide strong evidence \emph{against} premature black hole death, and therefore rule out much of the parameter space considered here. A caveat is that the James Webb Space Telescope is likely to only be capable of observing relatively late $z\lesssim 7.5$ PISN~\cite{2018MNRAS.479.2202H}, while premature black hole death would presumably be most efficient at earlier times, in pristine host halos. 

In summary, we have presented here a mechanism whereby very weakly-/non-annihilating dark matter could have been captured by Population III stars, leading to their premature death as black holes. The dynamics of capture, collapse and black hole growth have been discussed in great detail. The impact of this premature death on the epoch of reionisation is discussed in general terms, requiring further detailed study. We also note that other signatures, such as gravitational waves, or PISN, could be affected. These also deserve further study. Premature black hole death as a result of dark matter capture could be probed by any or all of these three signatures in the near future. The sensitivity could span almost 20 orders of magnitude in dark matter mass, and almost 40 orders of magnitude in interaction cross-section.

\section*{Acknowledgements}
We thank Anirban Das, Patrick Eggenberger, Raghuveer Garani, Shirley Li, Georges Meynet, Toby Opferkuch, Filippo Sala, Philip Schuster and Kevin Zhou for fruitful discussions. SARE was supported by SNF Ambizione grant PZ00P2\_193322, \emph{New frontiers from sub-eV to super-TeV}, in the final stages of this study. The authors would like to thank the Galileo Galilei Institute for Theoretical Physics (GGI, Florence), for hospitality during the research program on “New Physics from The Sky”.

\bibliographystyle{jhep}
\bibliography{BHAccretion}

\end{document}